\def\qquad{\hspace{30pt}}
\def\0{\newline}
\def\1{\newline}
\def\2{\par\noindent\newline}

\def\loq{,\kern-0.080em,\kern+0.05em}

\def\bfloq{{\bf,\kern-0.06em,\kern+0.05em}}

\def\footloq{,\kern-0.07em,\kern+0.03em}

\def\quabla{{\raise.7ex\hbox{\boxed{{}}}}}

\def\be{\begin{equation}}
\def\ee{\end{equation}}
\def\bea{\begin{eqnarray}}
\def\eea{\end{eqnarray}}
\def\bild#1#2#3{
\begin{figure}[hbt]
\epsfxsize=#3cm
\begin{center}
\leavevmode
\epsffile{#1.eps}
\end{center}
\caption[]{#2}
\label{#1}
\end{figure}
\vspace*{0.5cm}
}
 \documentstyle[12pt,epsf]{article}            
\title{%
{\Large\bf
    \vspace*{-50pt}
Robustness of the quantum Hall effect,
    \\[-08pt]                   
sample size versus sample topology,
    \\[-08pt]                   
and quality control management of
    \\[-10pt]                   
III-V molecular beam epitaxy}
     \\[20pt]  
       }
\author{%
{\sc Ralf D.\ Tscheuschner}%
\thanks{{\it permanent e-mail address:\/} ralfd@provi.de}%
, {\sc Sascha Hoch}, {\sc Eva Leschinsky},
     \\[-05pt]
{\sc Cedrik Meier}, {\sc Sabine Theis}, and {\sc Andreas D.\ Wieck}
     \\[20pt]  
Angewandte Festk\"orperphysik
\\[-05pt]
Ruhr-Universit\"at Bochum
\\[-05pt]
Universit\"atsstra\ss e 150
\\[-05pt]
D-44780 Bochum
\\[-05pt]
Federal Republic of Germany
\\[20pt]
}
\begin{document}
\maketitle
\begin{abstract}
We measure the IQHE
on macroscopic (1.5\,cm\,$\times$\,1.5\,cm)
\lq\lq quick and dirty\rq\rq\ prepared
III-V heterostucture samples with
{\sc van der Pauw} and modified {\sc Corbino}
geometries at 1.3 K.
We compare our results with 
(i) data taken on smaller specimens,
among them samples with a standard Hall bar geometry, 
(ii) results of our numerical analysis taking inhomogenities
of the 2DEG into account.
Our main finding is a confirmation of the expected
robustness of the IQHE
which favours the development of wide plateaux for
small filling factors and very large samples sizes
(here with areas 10\,000 times larger
than in standard arrangements).
\end{abstract}
\pagestyle{myheadings}
\pagenumbering{arabic}
\newpage  
\tableofcontents 
%
\vfill\pagebreak
\noindent%
%
%
%
%
\section{\hspace{5mm}Introduction and theoretical perspective}
A remarkable fact is that the inverse {\sc von\,Klitzing} constant
[1]
\begin{equation}
\frac{1}{R_{vK}} =
\frac{e^2}{h}    =
\frac{e}{h/e}
\end{equation}
is nothing but a ratio between an elementary electric and 
an elementary magnetic quantity, namely the elementary
electronic charge and the {\sc London} magnetic flux quantum.%
\footnote{%
\lq\lq {\sc London} magnetic flux quantum $h/e$\rq\rq\
as opposed to the \lq\lq BCS magnetic flux quantum $h/2e$\rq\rq\
reflecting the fact that the electrons,
from which the superconducting ground state
is built, are paired.} 
This is not unlike the expression for the
vacuum impedance
\begin{equation}
Z_0=\sqrt{\frac{\mu_0}{\varepsilon_0}},
\end{equation}
the fundamental quantity of r.f.\ technology.
In fact, the fundamental constant of quantum electrodynamics,
the {\sc Sommerfeld} fine structure constant, is given by the
ratio
(in MKSA units)
\begin{equation}
\alpha = \frac{Z_0}{2R_{vK}}
       = \frac{\sqrt{\mu_0/\varepsilon_0}}{2\,h/e^2}
       = \frac{e^2}{4\pi\varepsilon_0\,\hbar c}
       = \frac{e^2\mu_0c}{4\pi\hbar}.
\end{equation}
More strikingly, the ratio $g/e$ between the charge of a
hypothetical {\sc Dirac} magnetic monopole
[2]
and an electron charge is supposed to be
\begin{equation}
\frac{g}{e} = \frac{Z_0}{2\alpha}
            = R_{vK},
\end{equation}
since the {\sc Dirac} quantization condition
for a configuration consisting of
an elementary electric charge $e$
and an elementary magnetic charge $g$
reads (in MKSA)
\begin{equation}
\frac{1}{\hbar c}
\cdot
\frac{g}{\sqrt{4\pi\mu_0}}
\cdot
\frac{e}{\sqrt{4\pi\varepsilon_0}}
=
\frac{1}{2}.
\end{equation}
Notice, in terms of {\sc Gau\ss}ian quantities we have to write instead
\begin{equation}
\alpha = \frac{e^2}{\hbar c},
\phantom{xxxxx}
\frac{g}{e} = \frac{1}{2\alpha},
\phantom{xxxxx}
\frac{g \cdot e}{\hbar c} = \frac{1}{2}.
\end{equation}
\par
\lq\lq The {\sc von\,Klitzing} resistance is
       an universal ratio between a magnetic
       and an electric quantity.\rq\rq\
%
This statement suggests that the quantum {\sc Hall} effect
is a truly fundamental phenomenon of
quantum electrodynamics contrary to the popular belief
prevalent in semiconductor physics [3, 4, 5, 6, 7].%
\footnote{%
For example, {\sc Haldane} and {\sc Chen} strongly argue
against this view
[8].
They consider the material independence
of the QHE a strong evidence {\it against\/} electrodynamics effects.
They state that then in a physical realistic situation the effect
would depend on a material dependent \lq\lq effective fine structure
constant\rq\rq\ $\alpha'=(\mu/\varepsilon)^{1/2}/2R_{vK}$.
However, quantum physical quantization rules always
manifest themselves in terms of
bare (microscopic) quantities. One prominent example is the
quantization of circulation in neutral superfluids which
only depend on the bare mass (of the helium atoms for example)
in spite of the strong renormalization effects.
(Thanks one extended to
Professor {\sc Nils Schopohl} for reminding us of this point.)}
In particular, it implies that the
transversal conductivity plateaus appearing
at integer (resp.\ odd rational) multiples of
$e^2/h$ reflect a macroscopic quantum state
exhibiting a certain range of rigidity against a variation
of external parameters such as the strength of
the magnetic field or the density of charge carriers.
However, unlike BCS superconductivity
(including high-$T_C$ superconductivity), here we do
not encounter a Bose-like condensate made up from 
paired electric charges, i.e.\ the electrons (or holes),
but, obviously, a Bose-like condensate made up from
flux-charge composites%
\footnote{%
This interpretation is based on the composite boson model
by {\sc Zhang}, {\sc Hanson}, and {\sc Kivelson} for the fractional
effect, but it works for the integer one as well.
(For a theoretical framework providing a {\it unified\/} description
of the integer and fractional quantum {\sc Hall}
effects the reader is refered to {\sc Jain}s work
[9].)}
[9, 10, 11, 12].
Let us briefly recall the essence of the argument.
\par
The state characterized by a filling factor $\nu=1$
may be regarded as an assembly of bound states, each made up
from a point-like electric charge $e$ and an infinite thin
magnetic solenoid carrying a flux quantum $h/e$.
The cumulative
{\sc Aharonov}-{\sc Bohm}-{\sc Aharonov}-{\sc Casher} (ABAC)
[13, 14]
phase for adiabatically looping one bound state around
another is equal to
\begin{eqnarray}
&&
\mbox{{\it q.m.\,phase shift\/}}\,(\mbox{{\rm charge around vortex}})
\cdot
\mbox{{\it q.m.\,phase shift\/}}\,(\mbox{{\rm vortex around charge}})
\nonumber\\
&&
\phantom{123}
=
\exp\,\left\{\frac{i}{\hbar}\,   e\,(h/e) \right\}
\cdot
\exp\,\left\{\frac{i}{\hbar}\,   (h/e)\,e \right\} =
\exp\,\left\{\frac{i}{\hbar}\,2\,e\,(h/e) \right\} = 
\exp\, i 4\pi
= 1,
\end{eqnarray}
where, as usual, we have set $\hbar=h/2\pi$.
\par  
A simple exchange of two bound states, interpreted as an exchange 
of indistinguishable particles in the sense of quantum mechanics,
is topologically equivalent to one half of the above operation,
such that the statistics parameter becomes
\begin{equation}
\exp\,i\theta =    
\exp\,\left\{\frac{i}{\hbar}\,e\,(h/e) \right\} =
\exp\, i 2\pi = 1,
\end{equation}
where we assumed that the constituents, the electric as well 
as the magnetic ones, are all bosons.
It follows immediately that the composites are bosons at least
in a long-distance limit.
\par
However, if we try to incorporate these objects
into an action principle, i.e.\ into a Lagrangian framework,
things change dramatically: The composites become fermions, or,
if they are built from fermionic electric charges (as electrons are)
and bosonic flux lines, they transmute to bosonic charge-flux 
composites. Let us explain how this happens.
The Lagrangian function for an assembly of electrically
charged point-like particle in an external electromagnetic
potential is given by
%
\begin{equation}
L = L_{kin} - V({\bf x}_\alpha)
           + \sum_\alpha {\dot {\bf x}}_\alpha \cdot {\bf A}({\bf x}_\alpha)
  = L_{kin} -
           \sum_\alpha \frac{dx_\mu^\alpha}{dt} A^\mu,
\end{equation}      
with $x_\mu$ and $A_\mu$ being
the space-time coordinate and the
electromagnetic 4-potential, respectively.
The attachment of a flux line $\Phi$
to an electric charge may $Q$ be viewed
as an additional constraint, say
\begin{equation}
Q \propto \Phi,
\end{equation}
such that for the elementary quanta the relation
\begin{equation}
e = \frac{1}{R_{vk}} \cdot \frac{h}{e}.
\end{equation} 
is fulfilled.
In an oversimplified language of mathematical physics,
where we set $\hbar=h/2\pi=1$ as well as $e=1$,
the celebrated {\sc von\,Klitzing} resistance is simply $2\pi$,
and the constraint has to be rewritten as
\begin{equation}
Q = \frac{1}{2\pi}   \cdot \Phi.
\end{equation}
This is a poor man's form%
\footnote{freespoken after a joke by {\sc P.W.\ Anderson}
          on the renormalization group analysis of
          the {\sc Kondo} problem.}
of the {\sc Chern}-{\sc Simons} relation
[10, 11, 12].
Expressed in terms of the associated densities,
e.g.\ for one pointlike composite at the origin,  
\begin{eqnarray}
\varrho_{2D} (x,y) &=&         \delta(x)\delta(y) \\
B_z (x,y)          &=& 2\pi    \delta(x)\delta(y)
\end{eqnarray}  
we have
\begin{eqnarray}
\int \varrho_{2D} \, d^2 x &=& \frac{1}{2\pi} \, \int B_z d^2 x \nonumber  \\
                           &=& \frac{1}{2\pi} \, \int {\bf rot}\,{\bf A}\, d^2 x
                          \;=\;\frac{1}{2\pi} \, \oint {\bf A} d{\bf l}.
\end{eqnarray}                                 
Using a relativistic notation, which is of course the natural choice
in a context of a problem involving classical electrodynamics, we rewrite
this as
\begin{eqnarray}
j^0 &=& \frac{1}{2\pi} F^{12}                                   \nonumber \\
    &=& \frac{1}{2\pi} (\partial^1 A^2 - \partial^2 A^1)        \nonumber \\
    &=& \frac{1}{4\pi} \, \varepsilon_{012} \, 
                              (\partial^1 A^2 - \partial^2 A^1),
\end{eqnarray}
which, taking {\sc Lorentz} invariance into account, may be generalized to
\begin{equation}
j^\varrho = \frac{1}{4\pi} \, \varepsilon_{\varrho\sigma\tau} \, 
                              (\partial^\sigma A^\tau - \partial^\tau A^\sigma).
\end{equation}    
Inserting this constraint into the Lagrangian we finally get
\begin{eqnarray}
L &=\;& L_0 - \sum_\alpha
              \frac{dx^\alpha_\mu}{dt} A^\mu +
               \frac{1}{4\pi} \int d^2 x \, \varepsilon^{\mu\nu\varrho}
               A_\mu \partial_\nu A_\varrho                      \nonumber\\
         &=: & L_0 -
                              \int d^2 x \, j_\mu A^\mu +
               \frac{1}{4\pi} \int d^2 x \, \varepsilon^{\mu\nu\varrho}
               A_\mu \partial_\nu A_\varrho                      \nonumber\\
         &=: & L_0 -
               \left(\,
                              \int d^2 x \, j_\mu^{particles} -
               \frac{1}{4\pi} \int d^2 x \, \varepsilon^{\mu\nu\varrho}
               \partial_\nu A_\varrho
               \,\right) \,A^\mu                                 \nonumber\\
         &=: & L_0 -
               \int d^2 x \,
               (j_\mu^{particles} - j_\mu^{field}) \,A^\mu       \nonumber\\
         &=: & L_0 -
               \int d^2 x \,
               j_\mu^{total}\,A^\mu.
\end{eqnarray}
What now seems to come as a suprise is that the quantum mechanical
statistics parameter $\theta$ is exactly the fourth part of the denominator
of the topological constraint term, i.e.\ $\pi$. This can be verified
with help of functional integral techniques: Integrating out the
electromagnetic vector potential $A_\mu$ we get an effective non-local
action bilinear in the currents $j_\mu^{particles}$.
Calculations show that a two-particle-exchange trajectory
gives rise  to the correct phase factor.
The naive picture is consolidated if
we redefine the true electric charge of the charge-flux composite
as
\begin{equation}
Q_{true} = \int d^2 x\, j_0^{total} =
           \int d^2 x \left(j_0^{particles} - \frac{1}{4\pi}
           \varepsilon_{0\nu\lambda} F^{\nu\lambda} \right) = \frac{1}{2} \, Q,
\end{equation}
%
yielding the correct statistics phase factor
even in the ABAC inspired picture.%
\footnote{%
This point is missed in most popular treatments
on two-dimensional statistics (e.g.\ anyons).
The readers are often confused about
the very origin of statistics transmutation.
As shown, the flux-line pierced electron picture
has to be supplemented by a renormalization
of the effective electric charge of the composite
[15].}
Thus, if the picture is true,
a prerequesite for building the macroscopic Bose-condensed QHE state,
is the validity of the {\sc Chern}-{\sc Simons} dynamics, a fact 
emphasized by {\sc Fr\"ohlich}, {\sc Balachandran}, and others
[16, 17, 18].
Recently, {\sc Ghabhoussi} claimed the fundamental
validity of the {\sc Chern}-{\sc Simons} Lagrangian
for the {\it integral\/} quantum {\sc Hall} effect
[18].
However, the latter is a postulate,
at best comparable to the {\sc London} theory of superconductivity.
The fundamental problem is to find a microscopic justification
of this. 
\par
As early as 1984 {\sc Levine}, {\sc Libby}, and {\sc Pruisken}
[19]
as well as {\sc Pruisken} himself
[20]
decribed the integral quantum {\sc Hall} effect in the language
of a $\sigma$ model with a topological term, in which the
longitudinal and transversal components of the conductivity tensor,
$\sigma_{xx}$ and $\sigma_{xy}$, respectively,
play the role of coupling constants.
In an appropriate quantum field theoretical treatment, these are subject
to renormalization expressed in terms of a two-parameter scaling analysis
as shown in the pioneering work by {\sc Khmel'nitzkii}
[21].
{\sc Hall} conductivity plateaus correspond to vanishing
{\sc Callan}-{\sc Symanzik} $\beta$-functions, those points,
at which the quantized $\sigma$ model exhibits its conformal invariance.%
\footnote{%
The foliated phase structure of quantum field theories with
topological terms has been known for some time in the
high-energy physics community, see e.g.\
[22, 23].
In particular, Figs.\ 1 and 2
of
Ref.\ 23
anticipate the phase structure
of the full quantum {\sc Hall} problem
ten years before it became clear
[12].
One of us (R.D.T.) is indebted to
{\sc R.L.\ Stuller} for this remark.}
In spite of its sophistication and beauty, we think that even this model,
as well as many other related approaches, are build on presuppositions,
which already contain the expected result. Up to now, we have no microscopic
theory of the IQHE in which the exact quantization appears as a result,
not as a hidden assumption. Moreover, the debate whether the integral
quantum {\sc Hall} effect is a direct consequence of the fundamental
laws of a dimensionally reduced quantum electrodynamics or genuinely
tied to certain subtleties of semiconductor physics still seems to be open.
\par
A theory of the quantum {\sc Hall} effect should not only explain
the exact quantization of the transversal conductivity but also
describe the exact shape of the curves, which are only step and
delta functions in the limit of zero temperature. Clearly, {\sc Landau}
levels are broadened by impurity scattering, but this effect alone
does not explain the shape of the longitudinal and transversal resistivity
curves. For the many different approaches to the problem the reader is
refered to
[3,4,5,6,7].
\par
Some time ago {\sc Chang} and {\sc Tsui}
[24]
observed that the derivative of the finite-tem\-pe\-ra\-ture
quantum {\sc Hall} resistance $\varrho_{yx}$ with respect to the
two-dimensional carrier density $n_{2D}$ exhibits a remarkable
similarity to the longitudinal resistivity $\varrho_{xx}$ to the extent
that one is almost directly proportional to the other, i.e.\
\begin{equation}
\frac{d\varrho_{yx}}{dn_{2D}} \approx - a \cdot \varrho_{xx}.
\end{equation}
Two significant deviations of this behaviour should be
mentioned: Firstly, the relation does no longer hold in
the classical regime $B\rightarrow 0$ and, secondly,
the spikes appearing in the derivative of the transversal
resistivity are smeared out in the longitudinal one.
\par
What is the reason for this apparently fundamental
relation between the two quantities? {\sc Chang} and
{\sc Tsui} speculate about a {\sc Kramers}-{\sc Kr\"onig}
type relation based on causality
[25].
The presence of a natural frequency scale $\omega_c$,
the cyclotron frequency, and the suggestive association
of the longitudinal and transversal resistivities as parts
of a generalized complex resistivity describing
a general type of a dielectric response phenomenon
in the sense of {\sc Keldysh} {\it et al.\/}
[26]
should give rise to this kind of dispersion relation.
We cannot expect, however, that the standard {\sc Kubo} formula
treatment does provide a background for this speculation
as claimed by {\sc Chang} and {\sc Tsui} in their 1985 paper
[24]
since it certainly does not contain all the information necessary
for a thorough treatment of the electromagnetic response
problem defined by the quantum {\sc Hall} setup
[27].
\par
The relation discovered by {\sc Chang} and {\sc Tsui}
can only directly verified in gated systems where we can
continuously control the two-dimensional charge carrier
density $n_{2D}$. Therefore, in the context of this paper,
it is interesting to reexpress the derivative relation in
terms of the resistivities and the magnetic field alone.
If one assumes that the filling factor
\begin{equation}
\nu = \frac{n_{2D}h}{eB}
\end{equation}
is the relevant variable, we may write
\begin{equation}
\frac{d\varrho_{yx}}{dn}
=
\frac{d\varrho_{yx}}{d\nu} \frac{d\nu}{dn_{2D}}
=
\frac{d\varrho_{yx}}{dB} \frac{dB}{d\nu} \frac{d\nu}{dn_{2D}}
=
- \frac{1}{n_{2D}} \frac{d\varrho_{yx}}{dB} \cdot B.
\end{equation}
Combining both equations we obtain
\begin{equation}
\frac{d\varrho_{yx}}{dB} \approx
   a \cdot \frac{n_{2D}}{B} \cdot \varrho_{xx}.
\end{equation}
Within a simple scaling model {\sc Vagner} and {\sc Pepper}
discuss some generalizations of this formula
[28].
Assumptions on the nature of the impurity scattering, on the spatial
variations of the transversal resistivities, on the strength
of the applied magnetic field etc.\ restrict the possible
values of the exponents of the general,
still phenomenological, formula
\begin{equation}
\frac{d\varrho_{yx}}{dB} \approx
   a \cdot n_{2D} \cdot B^s \cdot \varrho_{xx}^t.
\end{equation}
In the cases accessible in our experiments we have
$t=1$ with $s$ being a small positive or negative
number of order $10^{-2}$ for a negative or positive
slope of $\varrho_{xx}(B)$ at $B=0$, respectively.
\par
We will use these phenomenological formulas
as a basis of numerical simulations where we
average over a finite number of replicas with
a certain distribution of values simulating
inhomogenities of the two-dimensional
charge carrier density $n_{2D}$ and
external magnetic field $B$.
\par
Our stategy is to push quantum {\sc Hall} experiments to the extremes
- in a truly literal sense!
If this effect is really a macroscopic quantum effect, then it should
be as robust as superconductivity, where it is possible to create
situations where the macroscopic quantum state is extended over
a region of many kilometers. Furthermore such a state should
exhibit a robustness which allows a \lq\lq quick 'n' dirty\rq\rq\
preparation. Nevertheless it still should exhibit features uniquely
associated with the topology of boundary conditions. This
paper is intended to be a first step toward a realization of
this strategy, following a (more or less) crazy suggestion
by one of us (A.D.W.) to Professor {\sc von\,Klitzing}
some time ago, namely to investigate the
quantized {\sc Hall} effects on huge samples
[29].
\par
To summarize, there are three main reasons to do so:
\begin{enumerate}
\item
To study the general limits of {\it macroscopic\/} quantum coherence
attributed to the quantum {\sc Hall} phenomenon.
\item
To study the scaling laws in the {\it infrared\/}
(using the terminology of quantum field theory),
i.e.\ in the large scale regime.
\item
To study the influence of inhomogenities
of the charge carrier density, the mobility,
and the external magnetic field on the
{\it \lq\lq spectral smearing\rq\rq\/} of the QHE signal.
\end{enumerate}
The latter topic seems to be of great importance
in quality control management of III-V (GaAs)
molecular beam epitaxy (MBE); for a review on this topic see
[30].
If it is possible to
interpret a huge sample quantum {\sc Hall} curve in an appropriate
way we will have a technique to measure the quality,
say {\it electrical\/} homogenity,
of a full wafer in a purely electronic, non-destructible way.
\par
The remaining part of the paper is organized as follows:
In the next section we present some computer simulations
based on a simple phenomenological model. This enables us
to get a feeling about the influence of inhomogenities
on the shape of the quantum {\sc Hall} curve. In what
follows we briefly describe the experimental set-up
including the preparation of the samples. Finally,
we review the experimental results, try to interpret
them, and make some suggestions towards future research.
%
%
\section{\hspace{5mm}Computer simulations}
The classical formula for the {\sc Hall} resistance
in case of a {\sc Hall} bar geometry is given by
\begin{equation}
R^{cl}_H = \frac{B}{en_{2D}}
         = \left. \frac{h}{\left(
                                 {\displaystyle \frac{hn_{2D}}{eB}}
                           \right) e^2 }
           \right. .
\end{equation}
The quantum analog reads
\begin{equation}
R^{qu}_H = \frac{h}{\nu e^2}.
\end{equation}
Consequently, an idealized quantum {\sc Hall} curve,
in which the {\sc Hall} resistance $R^{ideal}_H(B)$
is understood as a function depending on the external
magnetic field $B$ is given by the assignment
\begin{equation}
\nu \longmapsto
\nu(B) = \left\{
   \begin{array}{lllll}
      {\rm int} \left( {\displaystyle \frac{hn_{2D}}{eB} + \frac{1}{2} }
                \right)
      &\phantom{1234}
      &\mbox{{\rm if}}
      &\phantom{1234}
      &{\displaystyle \frac{hn_{2D}}{eB}>\frac{1}{2} }
      \\
      {\displaystyle \frac{hn_{2D}}{eB} }
      &\phantom{1234}
      &\mbox{{\rm else}}
      &\phantom{1234}
      &
   \end{array}
        \right.
\end{equation}
The \lq\lq else\rq\rq\ condition guarantees that below $\nu=1$
the curve is classical as will be expected if the fractional
effect is absent. The case including spin degeneracy is modelled
in an analogous way (Figures 1-4).
\par
Averaging may be done additively (arithmetic mean),
multiplicatively (geometric\linebreak
mean), or reciprocal
additively (harmonic mean) by scattering the
parameters for the two-di\-men\-si\-o\-nal charge carrier
density and the external magnetic field around selected
fixed values.
In case of a different geometry than the {\sc Hall} bar one
an additional multiplicative factor $\gamma$ has to be included
\begin{equation}
R^{cl}_H = \frac{\gamma B}{en_{2D}}
         = \left. \frac{h}{\left(
                                 {\displaystyle \frac{hn_{2D}}{e\gamma B}}
                           \right) e^2 }
           \right.
\end{equation}
in such a way that the quantization
(which must not depend on geometry) stays intact:
\begin{equation}
R^{qu}_H = \frac{h}{\nu e^2}.
\end{equation}
Geometry factors may vary as well
and can be absorbed in a redefinition of the
two-dimensional carrier density and magnetic field, respectively,
introducing new effective quantities $n_{2D}^{ef\!f}$ and $B^{ef\!f}$.
This may be useful in comparing totally different kinds of specimens.
\par
The mathematically inequivalent averaging methods correspond
to different quantum {\sc Hall}-sample-network models such as arrangements
in series, parallel, or combinations thereof. However, in cases
of a sufficiently narrow distribution of values ($\leq 6\,\%$)
it does not matter which method we prefer since then the results
of all averaging methods nearly coincide.
\par
Including in this algorithm
the proposal by {\sc Vagner} and {\sc Pepper}
[28]
we get a fairly good simulation of realistic
{\sc Hall} curves which explicitely show up the
smoothing of the steps and exhibit an extremely
realistic behaviour of the logitudinal resistance.
This can be seen in Figures 5-10.
Clearly, our averaging model does not include
microscopic mechanisms such as quantum interference.
However, our diagrams can be used as a reference
in the study of experimental curves to get
knowledge about the large scale behaviour.
%
%
\section{\hspace{5mm}Experiments with topological trivial samples}
We grow a GaAs/Al$_{0.33}$Ga$_{0.67}$As modulation doped
heterostructure by molecular beam epitaxy on a semi-insulating
GaAs (100) substrate. It consists of a 2 $\mu$m nominally doped
GaAs buffer layer and a 23 nm undoped Al$_{0.33}$Ga$_{0.67}$As
spacer layer, followed by 50 nm of Si doped Al$_{0.33}$Ga$_{0.67}$As
and a 10 nm GaAs cap. The two-dimensional electron gas is localized
in a sheet within the GaAs buffer right at the interface to the
spacer.
\par
The measured values for the electron sheet density and
the electron mobility are
\begin{itemize}
\item
at room temperature
3.36\,$\times$\,10$^{11}$\,cm$^{-2}$ and 6\,477 cm$^2\,$V$^{-1}$s$^{-1}$
\item
at T=77 K in the dark
3.33\,$\times$\,10$^{11}$\,cm$^{-2}$ and 84\,600 cm$^2\,$V$^{-1}$s$^{-1}$
\item
at T=5 K in the dark
3.1\,$\times$\,10$^{11}$\,cm$^{-2}$ and 336\,000 cm$^2\,$V$^{-1}$s$^{-1}$
\end{itemize}
\par
After the growth process the wafer ($\#$7235) is cut into parts.
One sample is mesa-etched into a standard {\sc Hall}-bar geometry with a width
of 150 $\mu$m and a distance of 200 $\mu$m between ohmic contacts
which were made with an AuGe/Ni alloy.
Thus in this type of specimen
(called \lq\lq {\bf micro}\rq\rq)
we have an electric active area
of some 0.1 mm$^2$
depending on the patching.
The other ones are square-shaped
{\sc van\,der\,Pauw}-type
[31]
specimens
of 3\,$\times$\,3 mm$^2$ = 9 mm$^2$
(called \lq\lq {\bf milli}\rq\rq)
of 1.5\,cm\,$\times$\,1.5\,cm = 625 mm$^2$
(called \lq\lq {\bf centi}\rq\rq),
respectively.
Roughly speaking, this collection of samples
enables us to
study real-space scaling experimentally over
four orders of magnitude, which is a lot!
(The next two orders of magnitude  would require specimens of about
100\,000 mm$^2$ corresponding to full 300 mm wafers,
which are not available yet.)
The larger specimens were contacted \lq\lq quick 'n' dirty\rq\rq\
by alloying-in some indium at the corners and the inner edges,
respectively.
This was done under an nitrogen-hydrogen atmosphere.
The samples are mounted on a chip carrier and
the measurement was done in a standard way
using a home-made%
\footnote{by one of us (A.D.W.)}
metal cryostat used in experimental lab courses.
The arrangements of the contacts and the sample geometry
are shown in Figures 11-16.
\par
Experimental results are shown in
Figures 17-27 and Tables 1-3.
We observed an interesting aging effect, a drop
of the signal for the longitudinal resistivity
which disappeared after an experimental rest
(Figures 17-22).
The effect could be reproduced and was probably
attributed to the thermodynamics of the set-up,
which eventually caused one sample to break
(see Figure 14).
\par
As a main result of our investigations,
the scaling behaviour can be read off
from Figures 23-25 and, finally,
from Figure 26 and Table 1.
\par
In the literature, scaling mostly is discussed
not by considering the renormalization of the
sample size in real space but, rather, in terms
of its low temperature behaviour.
Essentially, the dependence of conductance on
temperature is equivalent to its dependence on
the sample size. In a very interesting experiment
{\sc H.-P.\ Wei} {\it et al.\/} performed such
a scaling analysis for the quantum {\sc Hall} problem
[33, 34].
Essentially, they observed the behaviour of the
derivative of the transversal {\sc Hall} resistance
with respect to the temperature, which diverges
algebraically with $T \rightarrow 0$. For the
critical exponent which characterizes this divergence
they obtain from measurements between 4.2 and 0.1 K
a universal value of 0.42.
\par
In any finite-temperature experiment an effective sample size
is determined by the {\sc Thouless} length which is a measure
of the mean free path for inelastic scattering.
But the temperature-size analogy rests on certain
assumptions which may be questioned in the very large
scale limit, such that, of course, it makes sense
to perform a real-space experiment in the laboratory.
By rescaling the magnetic field the curves
in Figure 26 are normalized in such a way
that they can be compared directly. In Table 1
we list some characteristic properties of the
curves, from which the reader may find
an appropriate phenomenological formula.
In terms of our numerical simulations
the samples corresponds to a family of replica
of an idealized reference sample with a variation of parameters
(i.e.\ the plateau width) within a few percent range
showing again the robustness of the effect.
\par
Two qualitative observations
should be underlined: Firstly, whereas the
higher plateaux ($\nu>4$) are smeared out,
the lower seems to become more pronounced and
stable for very large samples, secondly,
the lifting of the spin degenacy is worse for
huge specimens. That is probably due to the fact
that polarized domains have an characteristic
maximum size.
\par
Of course, our samples are not ideal.
But, in general,
it is very difficult to distinguish between
genuine effects and effects attributed to
additional imperfections introduced by
the specific production process
which may be afflicted with one or
another shortcoming in the MBE growth.
According to common wisdom
the quantum {\sc Hall} effect is a localization-delocalization
phenomenon due to a mild form of disorder and
therefore it is almost impossible to define an
idealized reference sample. Nevertheless it would
be useful to do this empirically by repeating these
experiments again and again and logging them.
%
%
\section{\hspace{5mm}Measurements on topological non-trivial samples}
Like the setup proposed by {\sc van\,der\,Pauw}
[31]
the technique utilizing a {\sc Corbino} disk (i.e.\ a ring shaped sample)
can be used to measure {\sc Hall} mobilities
in different types of semiconductors, see e.g.\
[35],
or the low-temperature parameters of a two-dimensional electron gas
[36].
\par
In the study of the {\it quantized\/} {\sc Hall} effect
the {\sc Corbino} technique enables us to study situations
in which the samples are contactless (with respect to the
injected current) and hence the currents are edgeless.
This is interesting since it provides us with information
on the physics in the bulk. The connection between
2+1 dimensional {\sc Chern}-{\sc Simons} quantum field
theory and 1+1 dimensional conformal field theory indicate,
from first principles of quantum physics and gauge theory
alone, that the edge and the bulk pictures
should not be seen as concurrent but rather as
complementary and, hence, compatible approaches, e.g.\
[37, 38].
Thus the analysis of the {\sc Corbino} topology definitely
completes our understanding of a two-dimensional
quantum-electrodynamical response phenomenon.
\par
In the edgeless case current is induced inductively either
by modulating the external magnetic field with an a.c.\ driven
solenoid
[39, 40, 41]
or by a capacitive coupling
[42, 43].
The independent injection of different currents into the different
connected components of the topologically disconnected boundary
in a ring geometry was also studied
[44].
{\sc Wolf} {\it et al.\/} produced window-shaped quantum {\sc Hall}
effect samples with contacts both on the inner and on the outer
edges. This allowed them to study potentials appearing on contacts
inside a sample but still lying on an edge and hence not decoupling
from the two-dimensional electron gas
[45].
Their experimental results indicate that under quantum {\sc Hall} effect
conditions there is no electron transfer between the inner and the
outer edge.
\par
In our experiments we performed conventional quantum {\sc Hall}
measurements on a very large {\sc van\,der\,Pauw}-{\sc Corbino}
hybrid geometry/topology. The sample is a square shaped device of
1.5 cm$^2$ $\times$ 1.5 cm$^2$ with a centered hole of 7 mm diameter.
The latter was milled out by putting the varnished specimen in a spinner
and applying a pen-like rod of wood with sandpaper glued on its bottom.
(As a rather unconventional method it is without any respect.)
Contacts were soldered in exact the same way as in the case
of the samples with trivial topology. Experimental results are shown
in Figure 27 and commented in Table 3.
\par
Whereas it is interesting in itself that even in the topological
non-trivial case everything works fine with this
\lq\lq quick 'n' dirty\rq\rq\ preparation we should
point the reader onto two interesting additional observations:
Firstly, the behavior of the slopes in the longitudinal resistivity
depending on the fact whether it is measured inside or outside
(curves \mbox{\boldmath$\gamma$} and \mbox{\boldmath$\delta$})
and, secondly, the onset of a second-derivative content if the current
is injected on the different boundary than on which the voltage
is measured (curve \mbox{\boldmath$\zeta$}).
However, these structures are not always as pronounced and vary.
%
%
\section{\hspace{5mm}Conclusion}
From the viewpoint of a professional technician
our experiments may look a little bit sportive
if not amateurish.
However, from the viewpoint of a theoretician
who is interested in first principles
our investigations definitely have the flavour
of {\it realized\/} gedankenexperiments
intimately touching the first principles
underlying our field of interest,
namely the mesoscopic realization of
a dimensionally reduced 2+1 dimensional
quantum electrodynamics.
\par
The main result of our work is:
Yes, the integral quantum {\sc Hall} effect
is indeed a {\it macroscopic\/}, extremely robust,
quantum phenomenon.
\par
How far can we go? Clearly, one should repeat
all measurements on still larger samples
(up to 300 mm wafers, if they are available,
placed in high-energy accelerator detector magnets),
at lower temperatures (down to mK range)
and on samples of higher genera in the language of
analytic function theory (i.e.\ with more holes).
Of course, it would
be useful to study the scaling laws
in more detail experimentally although the
essence is already captured
in Figure 26 and Table 1.
\par
If we add techniques like focused ion beam
lithography, in particular in-plane gated
set-ups
[46, 47]
we probably could observe
the topological transition from a {\sc Corbino} disk
to a {\sc Hall} bar or from a {\sc Corbino} to
a {\sc van\,der\,Pauw} geometry.
In other words, we could perform experiments
within a unique topology changing scenario
in mesoscopic physics,
as it was recently done with help of different
preparation methods
[48].
Last but not least we should mention the famous
fountain pressure experiments by {\sc Klass} {\it et al.\/}
which could give us additional relevant information
about the physics in these huge samples
[49].
%
%
\section{\hspace{5mm}Acknowledgements}
The authors would like to thank all the members of
Angewandte Festk\"orperphysik for experimental support.
{\sc Eva Leschinksy} (Witten) gratefully
acknowledges the kind hospitality of the Bochum group.
One of us (R.D.T.) is indebted to
{\sc Sabine Gargosch} and {\sc Martin Versen} (Bochum),
{\sc Farhad Ghaboussi} (Konstanz),
Pro\-fes\-sor {\sc Nils Scho\-pohl} (Bochum/T\"ubingen),
{\sc Hermann He\ss\-ling} and {\sc Larry Stuller} (both at DESY Hamburg)
for inspring discussions.
%
%
\newpage
%

%
%
\section{\hspace{5mm}References}
\begin{itemize}
%
%
\item[{[1]}]
{\sc K.\ von\,Klitzing}, {\sc G.\ Dorda}, and {\sc M.\ Pepper},
{\it New method for high-accuracy determination of the fine-structure
     constant based on quantized Hall resistance\/},
Phys.\ Rev.\ Lett.\ {\bf 45}, 494-497 (1980)
%
%
\item[{[2]}]
{\sc P.A.M.\ Dirac},
{\it Quantized singularities in the electromagnetic field\/},
Proc.\ Roy.\ Soc.\ {\bf A133}, 60-72 (1931)
%
%
\item[{[3]}]
{\sc R.E.\ Prange} and {\sc S.M.\ Girvin},
{\it The Quantum Hall Effect\/},
Springer-Verlag, Berlin 1987
\item[{[4]}]
{\sc M.\ Stone} ed.,
{\it Quantum Hall Effect\/},
World Scientific, Singapore 1992
\item[{[5]}]
{\sc M.\ Janssen}, {\sc O.\ Viehweger},
{\sc U.\ Fastenrath}, and {\sc J.\ Hajdu},
{\it Introduction to the Theory of the Integer Quantum Hall Effect\/},
Verlag Chemie, Weinheim 1994
\item[{[6]}]
{\sc T.\ Chakraborty} and {\sc P.\ Pietil\"ainen},
{\it The Quantum Hall Effects\/},
Springer-Verlag, Berlin 1995
\item[{[7]}]
{\sc K.\ Efetov},
{\it Supersymmetry in Disorder and Chaos\/},
Cambridge University Press 1997
%
%
\item[{[8]}]
{\sc F.D.M.\ Haldane} and {\sc L.\ Chen},
{\it Magnetic flux of \lq\lq vortices\rq\rq\ on the
    two-dimensional Hall surface\/},
Phys.\ Rev.\ Lett.\ {\bf 53}, 2591 (1984)
%
%
\item[{[9]}]
{\sc J.K.\ Jain},
{\it Theory of the fractional quantum Hall effect\/},
Phys.\ Rev.\ {\bf B41}, 7653-7665 (1990)
\item[{[10]}]
{\sc S.-C.\ Zhang}, {\sc T.H.\ Hanson}, and {\sc S.\ Kivelson},
{\it Effective-field-theory model for the fractional quantum Hall effect\/},
Phys.\ Rev.\ Lett.\ {\bf 62}, 82-85 (1989)
\item[{[11]}]
{\sc S.-C.\ Zhang},
{\it The Chern-Simons-Landau-Ginzburg theory of the
     fractional quantum Hall effect\/},
Int.\ J.\ Mod.\ Phys.\ {\sc B6}, 25-58 (1992)
\item[{[12]}]
{\sc S.\ Kivelson}, {\sc D.-H.\ Lee}, and {\sc S.-C.\ Zhang},
{\it Global phase diagramm in the quantum Hall effect\/},
Phys.\ Rev.\ {\bf B46}, 2223-2238 (1992)
%
%
\item[{[13]}]
{\sc Y.\ Aharonov} and {\sc D.\ Bohm},
{\it Significance of electromagnetic potentials
    in the quantum theory\/},
Phys.\ Rev.\ {\bf 115}, 485-491 (1959)
\item[{[14]}]
{\sc Y.\ Aharonov} and {\sc A.\ Casher},
{\it Topological quantum effects for neutral particles\/},
Phys.\ Rev.\ Lett.\ {\bf 53}, 319-321 (1984)
%
%
\item[{[15]}]
{\sc X.G.\ Wen} and {\sc A.\ Zee},
{\it On the possibility of a statistics-changing phase transition\/},
J.\ Phys.\ France {\bf 50}, 1623-1629 (1989)
%
%
\item[{[16]}]
{\sc J.\ Fr\"ohlich} and {\sc A.\ Zee},
{\it Large scale physics of the quantum Hall fluid\/},
Nucl.\ Phys.\ {\bf B364}, 517-540 (1991)
\item[{[17]}]
{\sc A.P.\ Balachandran},
{\it Chern-Simons dynamics and the quantum Hall effect\/},
Pre\-print Syracuse SU-4228-492 (1991).
Published in a volume in honor of Professor R.\ Vijayaraghavan.
\item[{[18]}]
{\sc F.\ Ghaboussi},
{\it Quantum theory of the Hall effect\/},
Int.\ J.\ Theor.\ Phys.\ {\bf 36}, 923-934 (1997)
%
%
\item[{[19]}]
{\sc H.\ Levine}, {\sc S.B.\ Libby}, and {\sc A.M.M.\ Pruisken},
{\it Theory of the quantized Hall effect I\/},
Nucl.\ Phys.\ {\bf B240} (FS12), 30-48 (1984);
{\it II\/},
Nucl.\ Phys.\ {\bf B240} (FS12), 49-70 (1984);
{\it III\/},
Nucl.\ Phys.\ {\bf B240} (FS12), 71-90 (1984)
\item[{[20]}]
{\sc A.M.M.\ Pruisken},
{\it On localization in the theory of the quantized Hall effect:
     a two dimensional realization of the theta vacuum\/},
Nucl.\ Phys.\ {\bf B235} (FS11), 277-298 (1984)
%
%
\item[{[21]}]
{\sc D.E.\ Khmel'nitzkii},
{\it Quantization of Hall conductivity\/},
JETP Lett.\ {\bf 38}, 552-556 (1983)
%
%
\item[{[22]}]
{\sc J.L.\ Cardy} and {\sc E.\ Rabinovici},
{\it Phase structure of Z(P) models
     in the presence of a theta parameter\/},
Nucl.\ Phys.\ {\bf B205} (FS5) 1-16 (1982)
\item[{[23]}]
{\sc J.L.\ Cardy} and {\sc E.\ Rabinovici},
{\it Duality and the theta parameter
     in abelian lattice models\/},
Nucl.\ Phys.\ {\bf B205} (FS5) 17-26 (1982)
%
%
\item[{[24]}]
{\sc A.M.\ Chang} and {\sc D.C.\ Tsui},
{\it Experimental observation of a striking similarity between
     quantum Hall transport coefficients\/},
Solid State Comm.\ {\bf 56}, 153-154 (1985)
\item[{[25]}]
{\sc G.D.\ Mahan},
{\it Many-Particle Physics\/},
Plenum Press, New York 1986
\item[{[26]}]
{\sc L.V.\ Keldysh}, {\sc D.A.\ Kirzhnitz}, and {\sc A.A.\ Maraduduin},
{\it The dielectric function of condensed systems\/},
North-Holland, Amsterdam 1989
\item[{[27]}]
{\sc N.\ Schopohl}, private communication.
\item[{[28]}]
{\sc I.D.\ Vagner} and {\sc M.\ Pepper},
{\it Similarity between quantum Hall transport coefficients\/},
Phys.\ Rev.\ {\bf 37}, 7147-7148 (1988)
%
%
\item[{[29]}]
{\sc K.\ von\,Klitzing}, private communication.
%
%
\item[{[30]}]
{\sc M.A.\ Hermann} and {\sc H.\ Sitter},
{\it Molecular Beam Epitaxy\/},
Springer-Verlag, Berlin (1989)
%
%
\item[{[31]}]
{\sc L.J.\ van\,der\,Pauw},
{\it A method of measuring specific resistivity and Hall effects
     of discs of arbitrary shape\/},
Philips Research Reports {\bf 13}, 1-9 (1958)
%
%
\item[{[32]}]
{\sc M.A.\ Hermann} and {\sc H.\ Sitter},
{\it Molecular Beam Epitaxy\/},
Springer-Verlag, Berlin (1989)
%
%
\item[{[33]}]
{\sc H.P.\ Wei}, {\sc D.C.\ Tsui},
{\sc M.A.\ Paalanen}, and {\sc A.M.M.\ Pruisken},
{\it Experiments on delocalization and universality in the integral
     quantum Hall effect\/},
Phys.\ Rev.\ Lett.\ {\bf 61}, 1294-1296 (1988)
\item[{[34]}]
{\sc H.P.\ Wei}, {\sc S.Y.\ Lin},
{\sc D.C.\ Tsui}, and {\sc A.M.M.\ Pruisken},
{\it Effect of long-range potential fluctuations on scaling in the
     integer quantum Hall effect\/},
Phys.\ Rev.\ {\bf 45}, 3926-3928 (1992)
%
%
\item[{[35]}]
{\sc G.P.\ Carver},
{\it A Corbino disk apparatus to measure Hall
     mobilities in amorphous semiconductors\/},
Rev.\ Scient.\ Instr.\ {\bf 43}, 1257-1263 (1972)
\item[{[36]}]
{\sc I.M.\ Templeton},
{\it A simple contactless method for evaluating the low-temperature
     parameters of a two-dimensional electron gas\/},
J.\ Appl.\ Phys.\ {\bf 62}, 4005-4007 (1987)
%
%
\item[{[37]}]
{\sc A.P.\ Balachandran}, {\sc L.\ Chandar}, and {\sc B.\ Sathiapalan},
{\it Duality and the fractional quantum Hall effect\/},
Nucl.\ Phys.\ {\bf B443}, 465-500 (1995)
\item[{[38]}]
{\sc A.P.\ Balachandran}, {\sc L.\ Chandar}, and {\sc B.\ Sathiapalan},
{\it Chern-Simons duality and the fractional quantum Hall effect\/},
Int.\ J.\ Mod.\ Phys.\ {\bf A11}, 3587-3608 (1996)
%
%
\item[{[39]}]
{\sc P.F.\ Fontein}, {\sc J.M.\ Lagemaat},
{\sc J.\ Wolter}, and {\sc J.P.\ Andre},
{\it Magnetic field modulation - a method for measuring the Hall
     conductance with a Corbino disc\/},
Semiconductor Science and Technology {\bf 3}, 915-918 (1988)
\item[{[40]}]
{\sc B.\ Jeanneret}, {\sc B.D.\ Hall},
{\sc H.-J.\ Buhlmann}, {\sc R.\ Houdre},
{\sc M.\ Ilegems}, {\sc B.\ Jeckelmann}, and {\sc U.\ Feller},
{\it Observation of the integer quantum Hall effect by magnetic coupling
     to a Corbino ring\/},
Phys.\ Rev.\ {\bf B51}, 9752-9756 (1995)
\item[{[41]}]
{\sc B.\ Jeanneret}, {\sc B.D.\ Hall},
{\sc B.\ Jeckelmann}, {\sc U.\ Feller},
{\sc H.-J.\ Buhlmann}, and {\sc M.\ Ilegems},
{\it AC measurements of edgeless currents in a Corbino ring in the
     quantum Hall regime\/},
Solid State Comm.\ {\bf 102}, 287-290 (1997)
\item[{[42]}]
{\sc C.L.\ Petersen} and {\sc O.P.\ Hansen},
{\it Two-dimensional electron gases in the quantum Hall regime: analysis
     of the circulating current in contactless Corbino geometry\/},
Solid State Comm.\ {\bf 98}, 947-950 (1996)
\item[{[43]}]
{\sc C.L.\ Petersen} and {\sc O.P.\ Hansen},
{\it The diagonal and off-diagonal AC conductivity of two-dimensional
     electron gases with contactless Corbino geometry in the
     quantum Hall regime\/},
J.\ Appl.\ Phys.\ {\bf 80}, 4479-4483 (1996)
\item[{[44]}]
{\sc R.G.\ Mani},
{\it Steady-state bulk current at high magnetic fields
     in Corbino-type\linebreak
     GaAs/AlGaAs heterostructure devices\/},
Europhys.\ Lett.\ {\bf 36}, 203-208 (1996)
%
%
\item[{[45]}]
{\sc H.\ Wolf}, {\sc G.\ Hein},
{\sc L.\ Bliek}, {\sc G.\ Weimann}, and {\sc W.\ Schlapp},
{\it Quantum Hall effect in devices with an inner boundary\/},
Semiconductor Science and Technology {\bf 5}, 1046-50 (1990)
%
%
\item[{[46]}]
{\sc R.J.\ Haug}, {\sc A.D.\ Wieck}, and {\sc K.\ von\,Klitzing},
{\it Magnetotransport properties of Hall-bar with
     focused-ion-beam written in-plane-gate\/},
Physica {\bf B184}, 192-196 (1993)
\item[{[47]}]
{\sc R.D.\ Tscheuschner} and {\sc A.D.\ Wieck},
{\it Quantum ballistic transport in in-plane-gate transistors
     showing onset of a novel ferromagnetic phase transition\/},
Superlattices and Microstructures {\bf 20}, 616-622 (1996)
\item[{[48]}]
{\sc A.S.\ Sachrajda}, {\sc Y.\ Feng},
{\sc R.P.\ Taylor}, {\sc R.\ Newbury},
{\sc P.T.\ Coleridge}, {\sc J.P.\ McCaffrey},
{\it The topological transition from a Corbino to Hall bar geometry\/},
Superlattices and Microstructures {\bf 20}, 651-656 (1996)
\item[{[49]}]
{\sc U.\ Klass}, {\sc W.\ Dietsche},
{\sc K.\ von\,Klitzing}, and {\sc K.\ Ploog},
{\it Fountain-pressure imaging of the dissipation in quantum-Hall
     experiments\/},
Physica {\bf B169}, 363-367 (1991)
%
%
\end{itemize}
\newpage
%

\section{\hspace{5mm}Figures: Numerical simulations}
%
\vspace*{\fill}
\bild{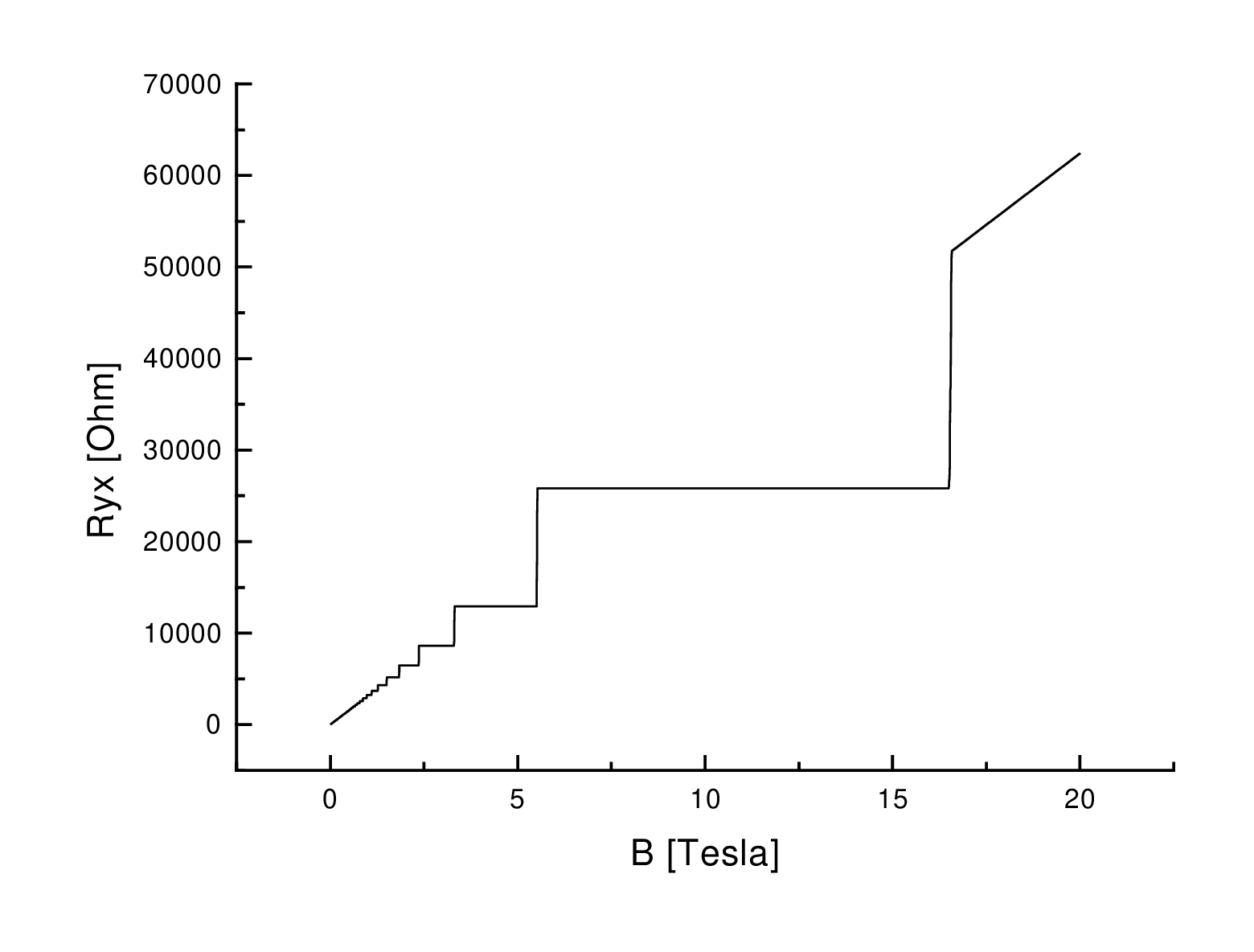}
     {Nearly ideal {\it transversal\/} quantized {\sc Hall} resistance
      (without spin degeneracy)
      down to filling factor $\nu=1$.
      (Computer simulation based on a phenomenological
      formula averaging over $N>50$ replicas within
      a charge carrier density variance
      $\sigma<0.1\,\%$ around $n_{2D}\approx 2.0 \cdot 10^{15}/{\rm m}^2$.)}
     {16}
\vspace*{\fill}
\eject\noindent%
\vspace*{\fill}
\bild{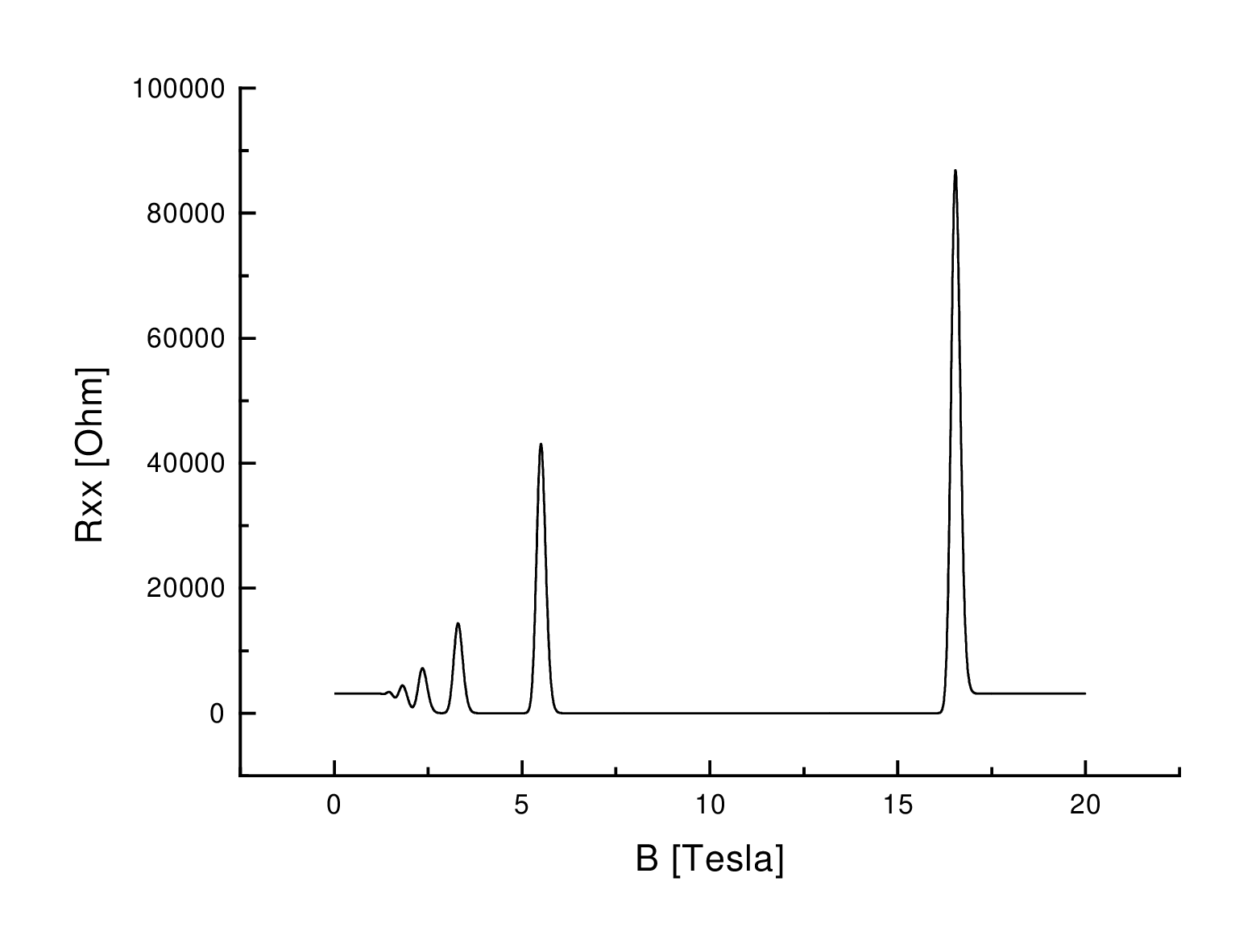}
     {Nearly ideal {\it longitudinal\/} quantized {\sc Hall} resistance
      (without spin degeneracy)
      down to filling factor $\nu=1$.
      (Computer simulation based on a phenomenological
      formula averaging over $N>50$ replicas within
      a charge carrier density variance
      $\sigma<0.1\,\%$ around $n_{2D}\approx 2.0 \cdot 10^{15}/{\rm m}^2$.)}
     {16}
\vspace*{\fill}
\eject\noindent%
%
%
\vspace*{\fill}
\bild{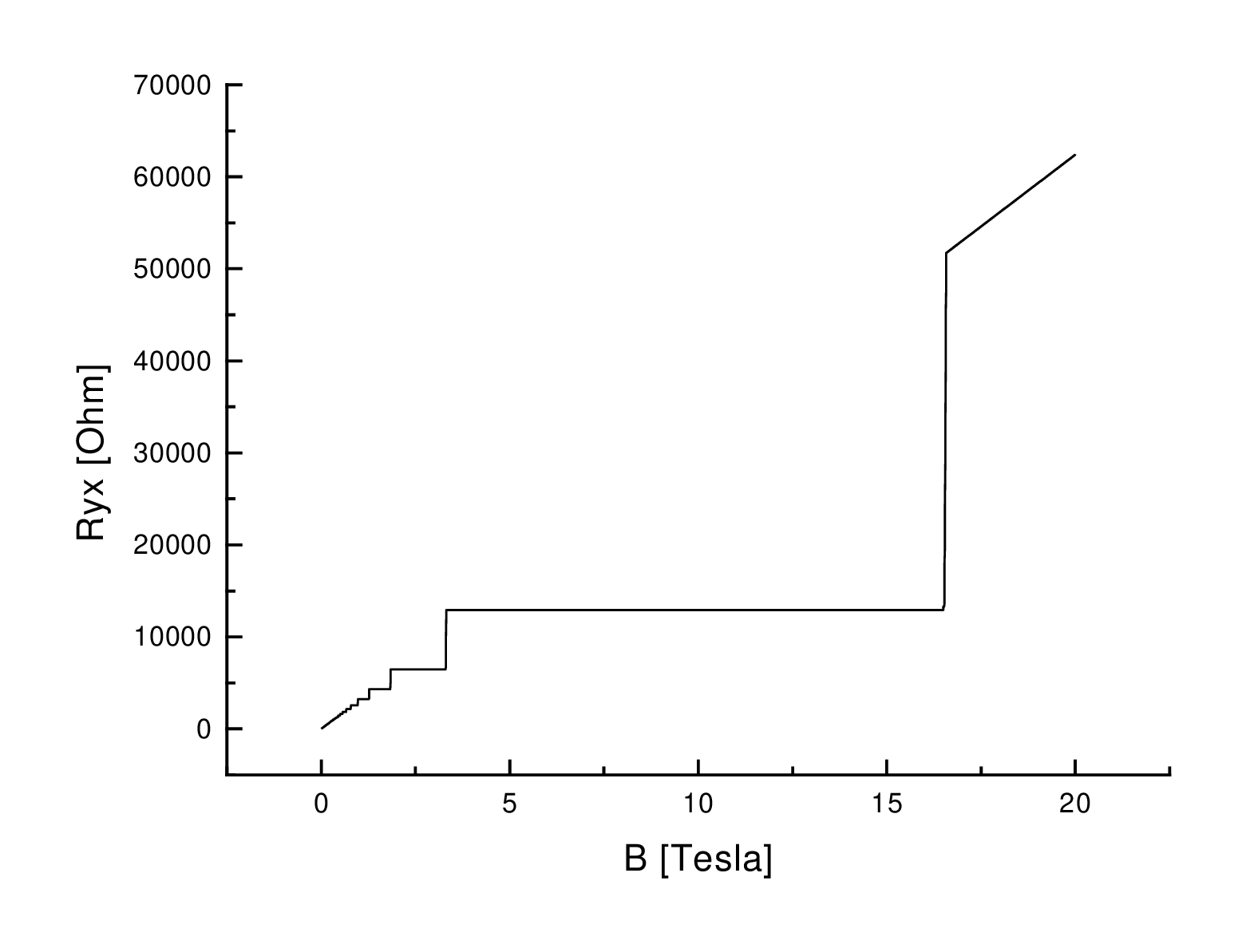}
     {Nearly ideal {\it transversal\/} quantized {\sc Hall} resistance
      {\it including\/} spin degeneracy
      down to filling factor $\nu=2$.
      (Computer simulation based on a phenomenological
      formula averaging over $N>50$ replicas within
      a charge carrier density variance
      $\sigma<0.1\,\%$ around $n_{2D}\approx 2.0 \cdot 10^{15}/{\rm m}^2$.)}
     {16}
\vspace*{\fill}
\eject\noindent%
\vspace*{\fill}
\bild{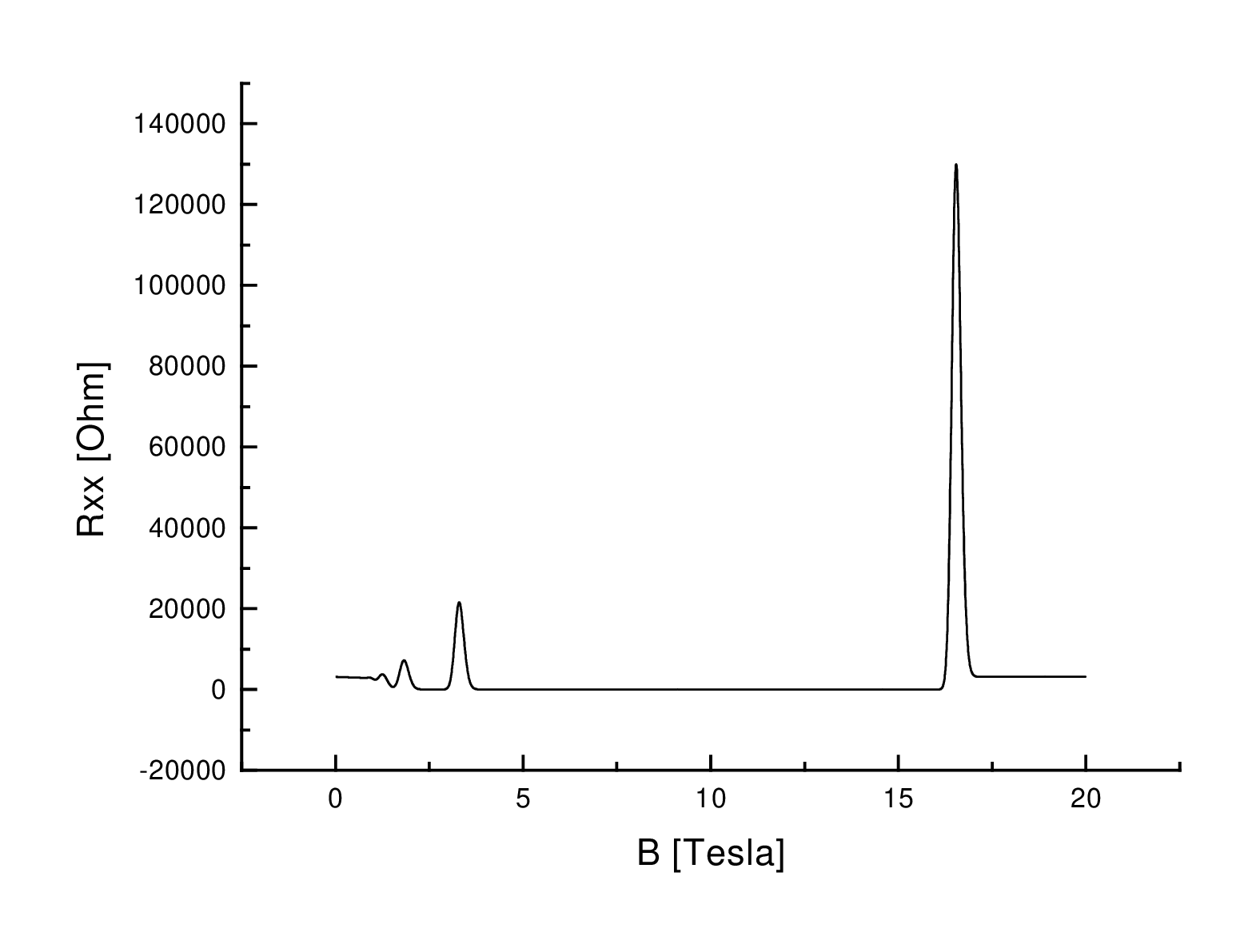}
     {Nearly ideal {\it longitudinal\/} quantized {\sc Hall} resistance
      {\it including\/} spin degeneracy
      down to filling factor $\nu=2$.
      (Computer simulation based on a phenomenological
      formula averaging over $N>50$ replicas within
      a charge carrier density variance
      $\sigma<0.1\,\%$ around $n_{2D}\approx 2.0 \cdot 10^{15}/{\rm m}^2$.)}
     {16}
\vspace*{\fill}
\eject\noindent%
%
%
\vspace*{\fill}
\bild{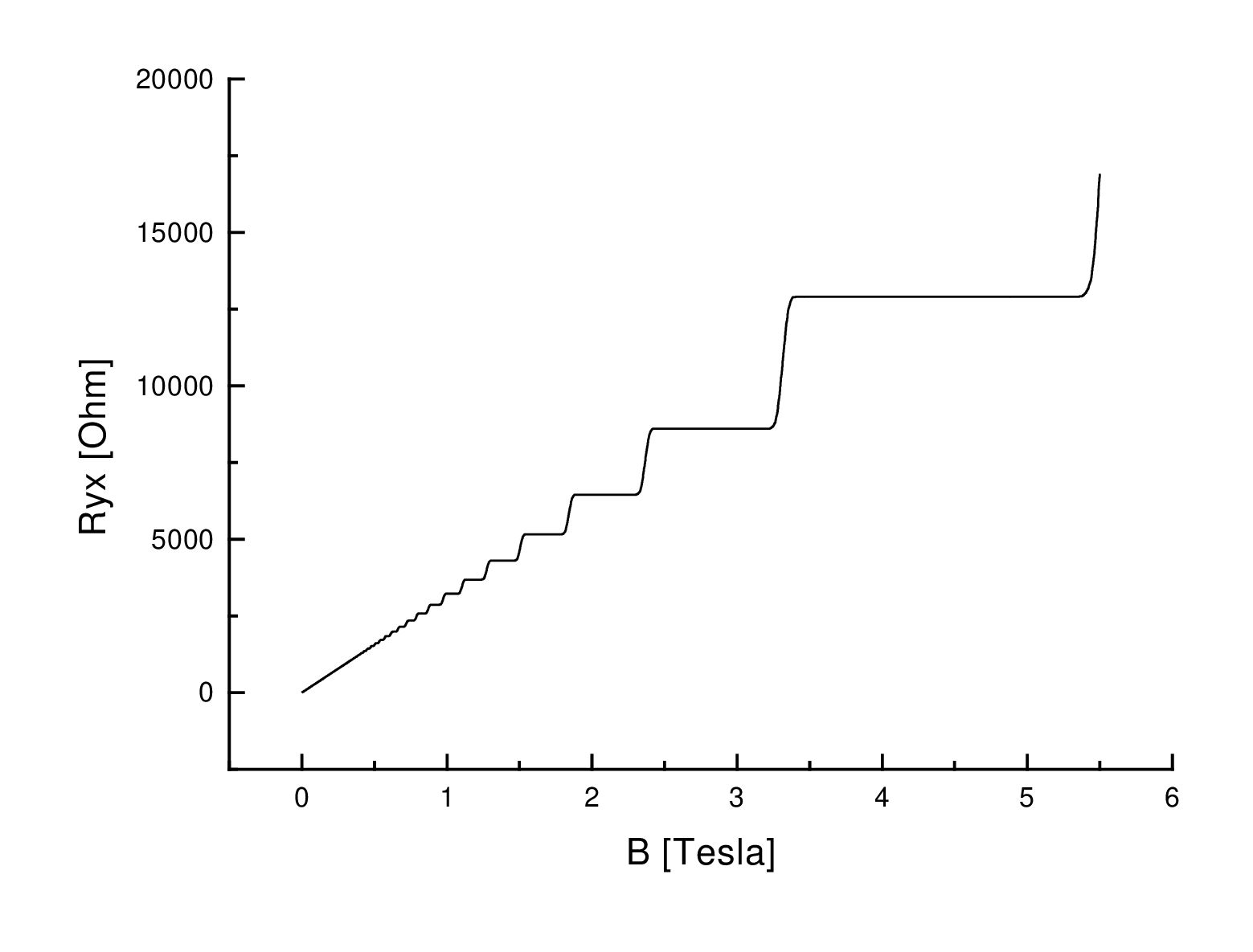}
     {{\it Transversal\/} quantized {\sc Hall} resistance
      (without spin degeneracy)
      down to filling factor $\nu=1$.
      (Computer simulation based on a phenomenological
      formula {\it geometrically\/} averaging over $N=1000$ replicas within
      a charge carrier density variance
      $\sigma=1\,\%$ around $n_{2D}\approx 2.0 \cdot 10^{15}/{\rm m}^2$.
      Inhomogenities of the external magnetic field may be modelled
      in an analogous fashion.)}
     {16}
\vspace*{\fill}
\eject\noindent%
\vspace*{\fill}
\bild{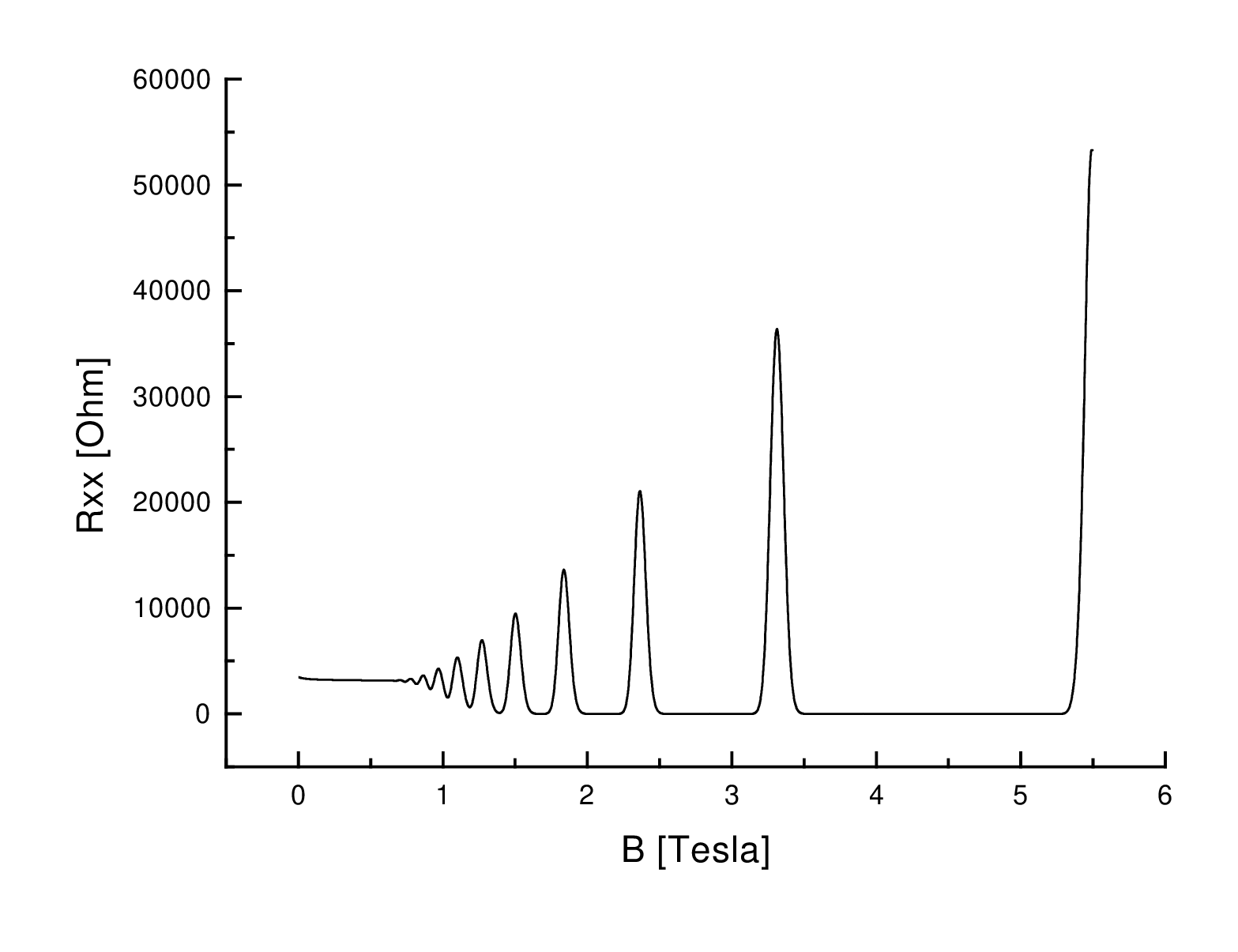}
     {{\it Longitudinal\/} quantized {\sc Hall} resistance
      (without spin degeneracy)
      down to filling factor $\nu=1$.
      (Computer simulation based on a phenomenological
      formula {\it geometrically\/} averaging over $N=1000$ replicas within
      a charge carrier density variance
      $\sigma=1\,\%$ around $n_{2D}\approx 2.0 \cdot 10^{15}/{\rm m}^2$.
      Inhomogenities of the external magnetic field may be modelled
      in an analogous fashion.)}
     {16}
\vspace*{\fill}
\eject\noindent%
%
%
\vspace*{\fill}
\bild{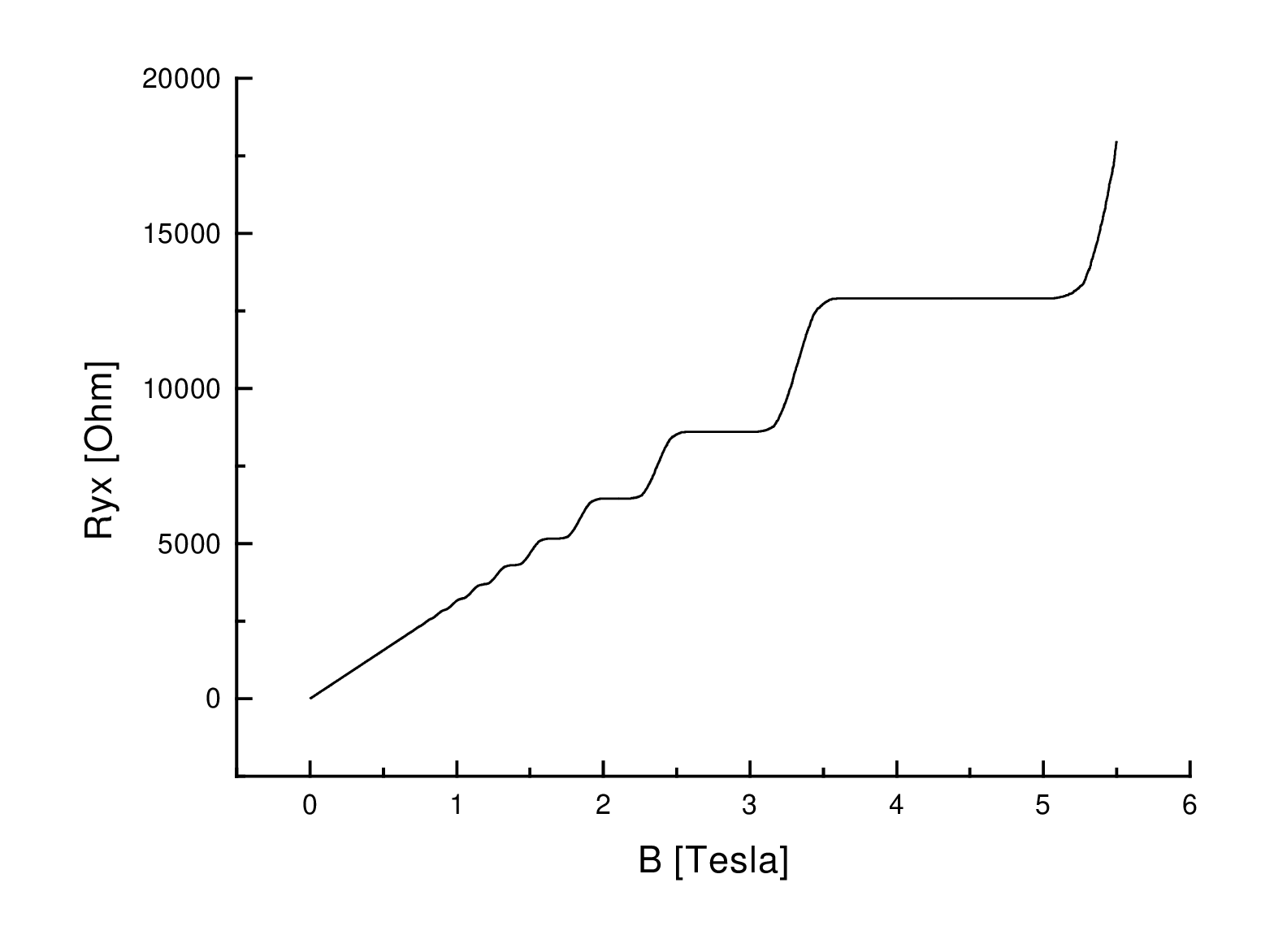}
     {{\it Transversal\/} quantized {\sc Hall} resistance
      (without spin degeneracy)
      down to filling factor $\nu=1$.
      (Computer simulation based on a phenomenological
      formula {\it geometrically\/} averaging over $N=1000$ replicas within
      a charge carrier density variance
      $\sigma=3\,\%$ around $n_{2D}\approx 2.0 \cdot 10^{15}/{\rm m}^2$.
      Inhomogenities of the external magnetic field may be modelled
      in an analogous fashion.)}
     {16}
\vspace*{\fill}
\eject\noindent%
\vspace*{\fill}
\bild{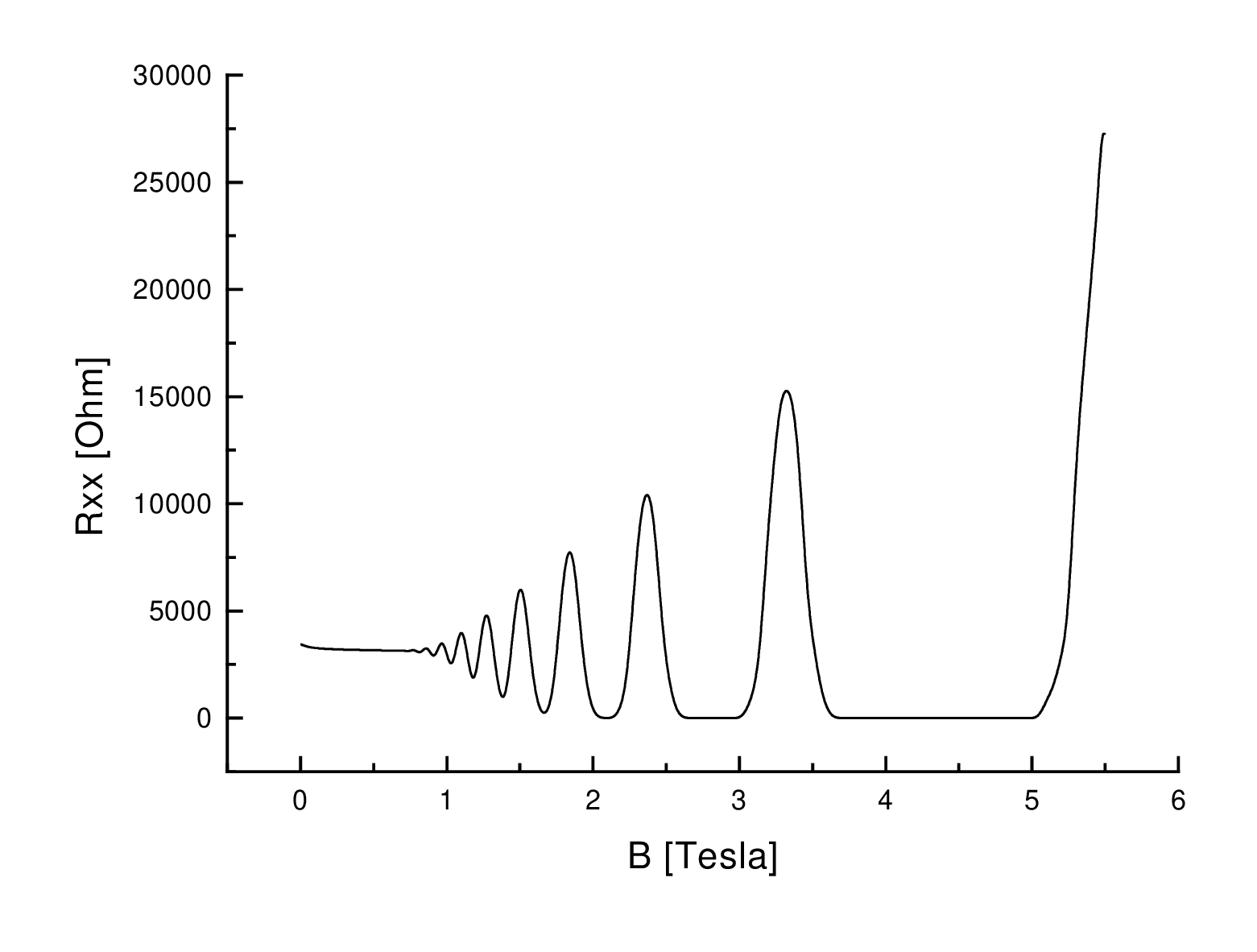}
     {{\it Longitudinal\/} quantized {\sc Hall} resistance
      (without spin degeneracy)
      down to filling factor $\nu=1$.
      (Computer simulation based on a phenomenological
      formula {\it geometrically\/} averaging over $N=1000$ replicas within
      a charge carrier density variance
      $\sigma=3\,\%$ around $n_{2D}\approx 2.0 \cdot 10^{15}/{\rm m}^2$.
      Inhomogenities of the external magnetic field may be modelled
      in an analogous fashion.)}
     {16}
\vspace*{\fill}
\eject\noindent%
%
%
\vspace*{\fill}
\bild{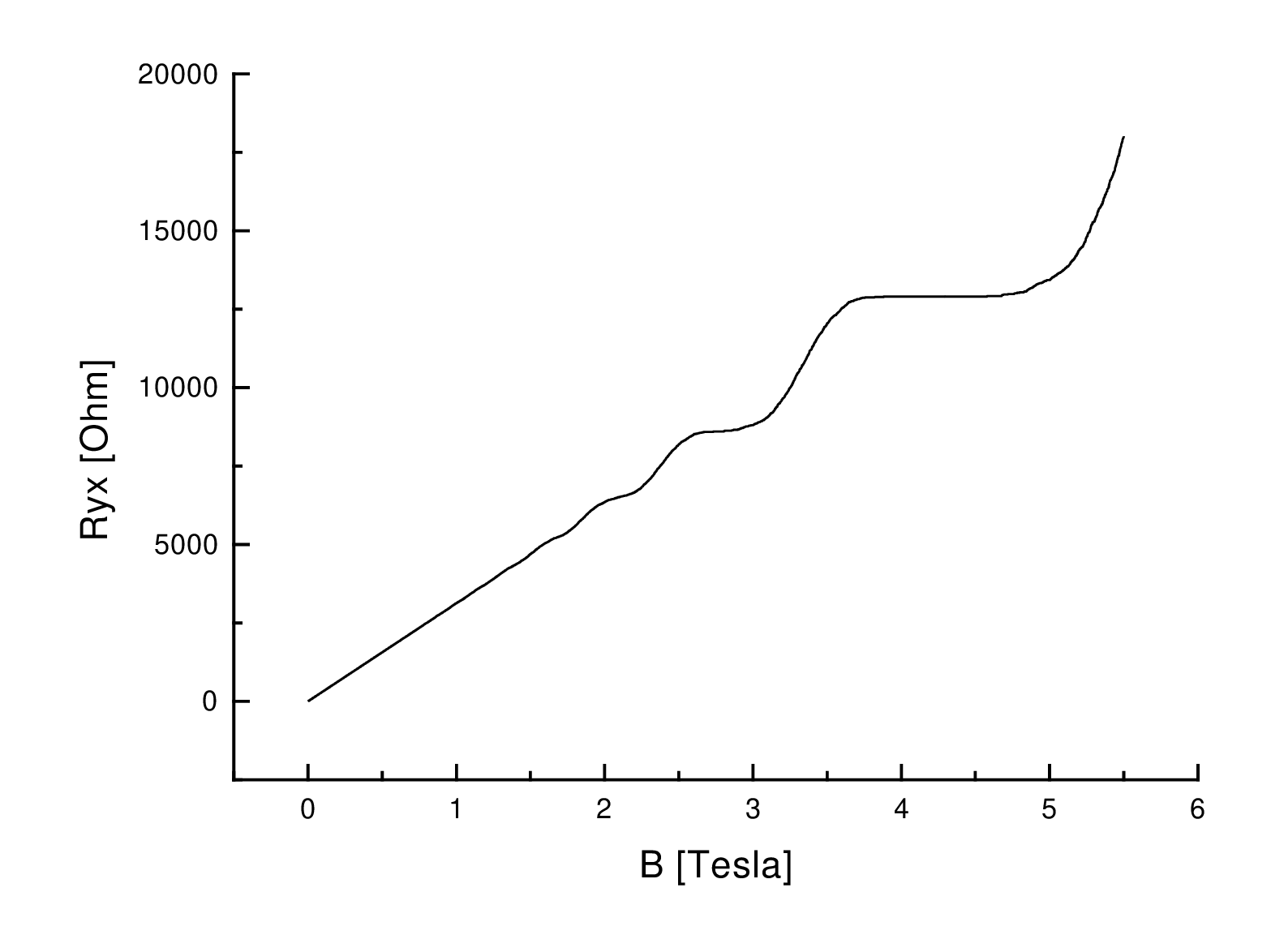}
     {{\it Transversal\/} quantized {\sc Hall} resistance
      (without spin degeneracy)
      down to filling factor $\nu=1$.
      (Computer simulation based on a phenomenological
      formula {\it geometrically\/} averaging over $N=1000$ replicas within
      a charge carrier density variance
      $\sigma=6\,\%$ around $n_{2D}\approx 2.0 \cdot 10^{15}/{\rm m}^2$.
      Inhomogenities of the external magnetic field may be modelled
      in an analogous fashion.)}
     {16}
\vspace*{\fill}
\eject\noindent%
\vspace*{\fill}
\bild{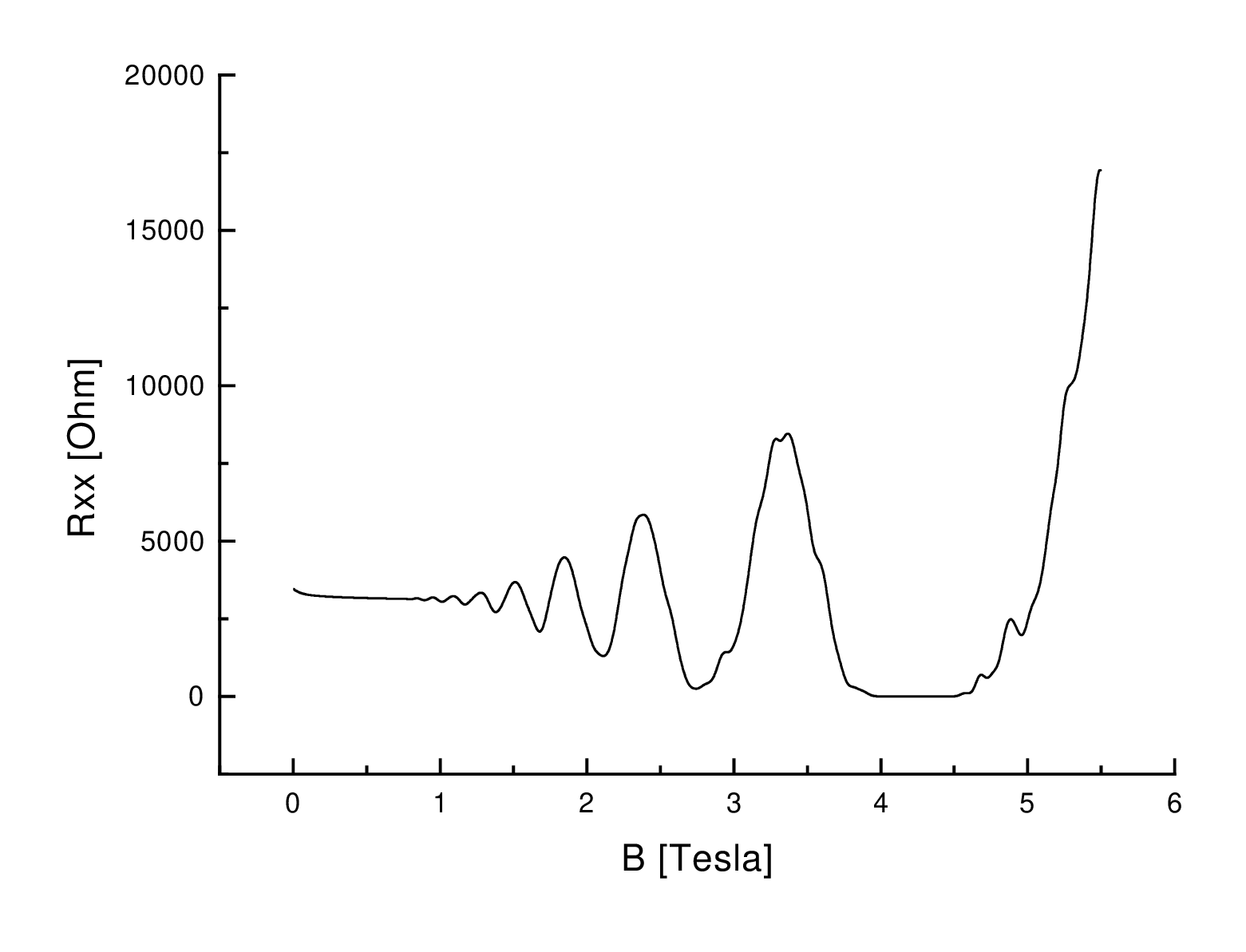}
     {{\it Longitudinal\/} quantized {\sc Hall} resistance
      (without spin degeneracy)
      down to filling factor $\nu=1$.
      (Computer simulation based on a phenomenological
      formula {\it geometrically\/} averaging over $N=1000$ replicas within
      a charge carrier density variance
      $\sigma=6\,\%$ around $n_{2D}\approx 2.0 \cdot 10^{15}/{\rm m}^2$.
      Inhomogenities of the external magnetic field may be modelled
      in an analogous fashion.)}
     {16}
\vspace*{\fill}
%
\newpage

\section{\hspace{5mm}Figures: Sample geometry and topology}
%
\vspace*{\fill}
\bild{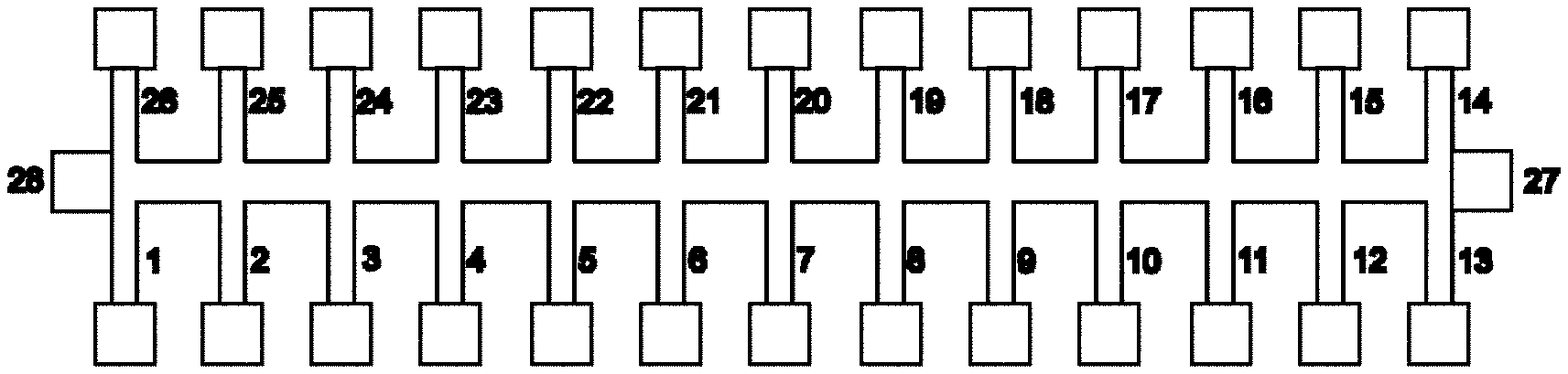}{{\sc Hall}-bar layout}{15}
\vspace*{\fill}
\eject\noindent%
\vspace*{\fill}
\bild{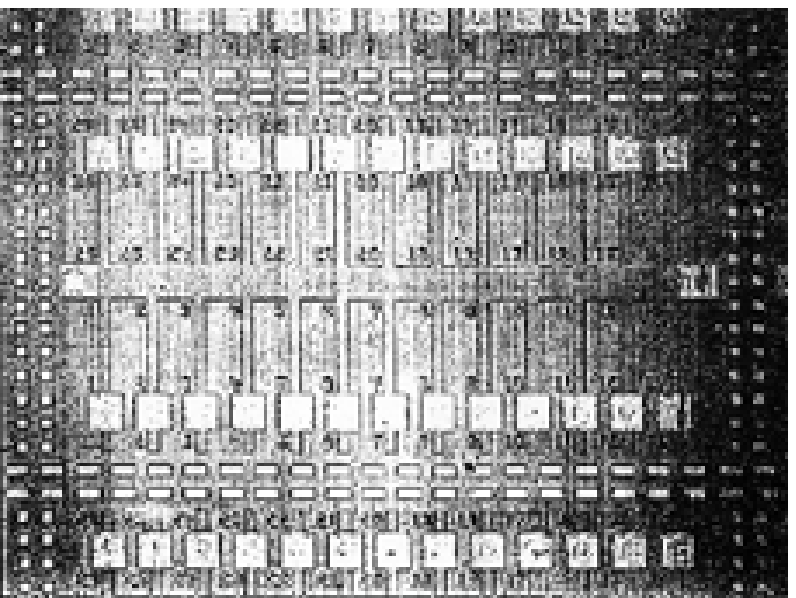}{Photograph of the {\sc Hall}-bar}{12}
\vspace*{\fill}
\eject\noindent%
\vspace*{\fill}
\bild{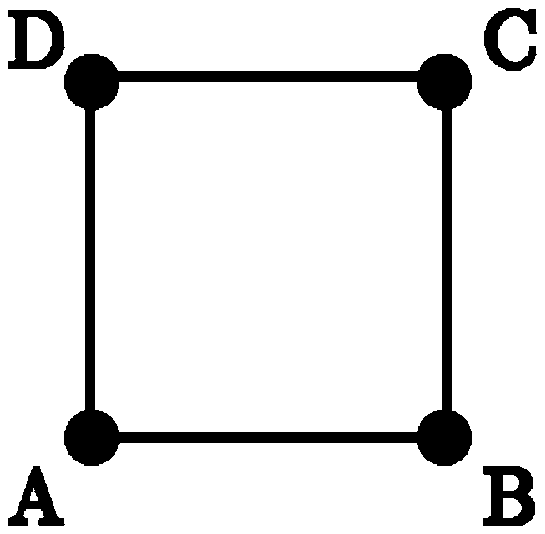}{{\sc van\,der\,Pauw}-type sample layout}{6}
\vspace*{\fill}
\eject\noindent%
\vspace*{\fill}
\bild{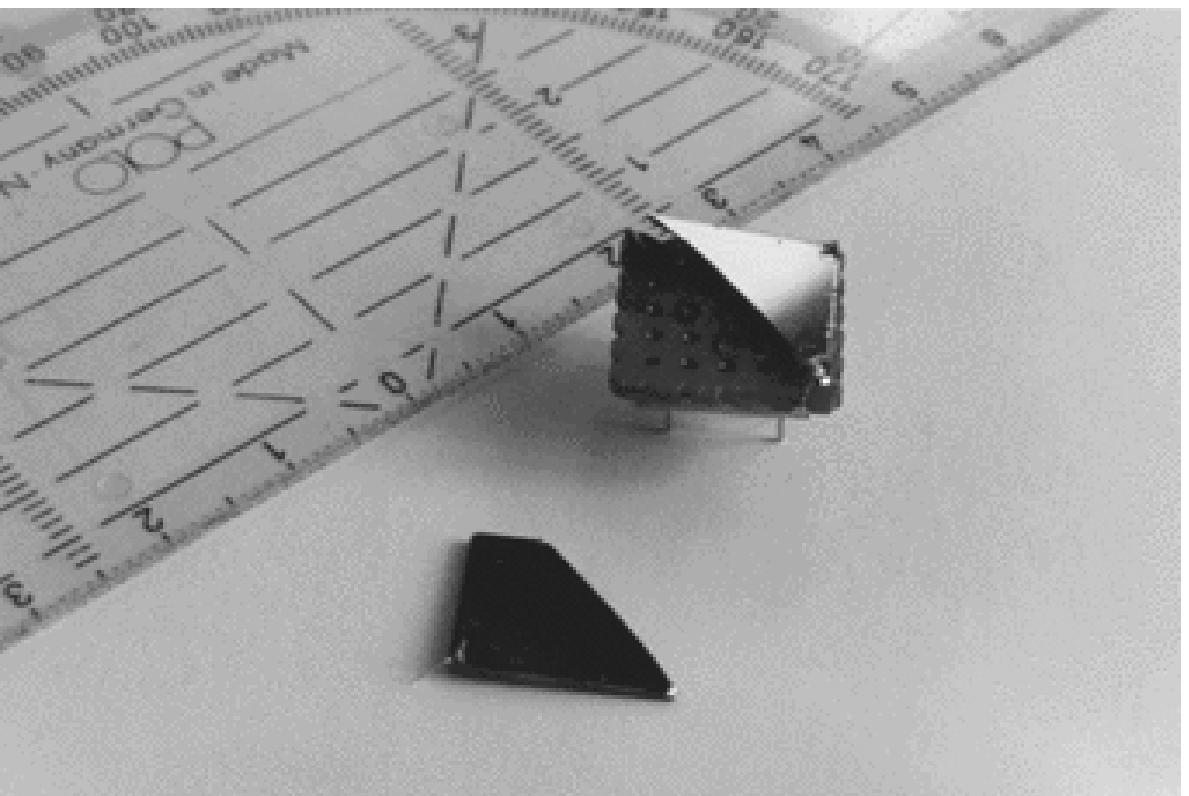}{Photograph of the (broken)
                {\sc van\,der\,Pauw}-type \lq\lq{\bf centi}\rq\rq\ sample}{12}
\vspace*{\fill}
\eject\noindent%
\vspace*{\fill}
\bild{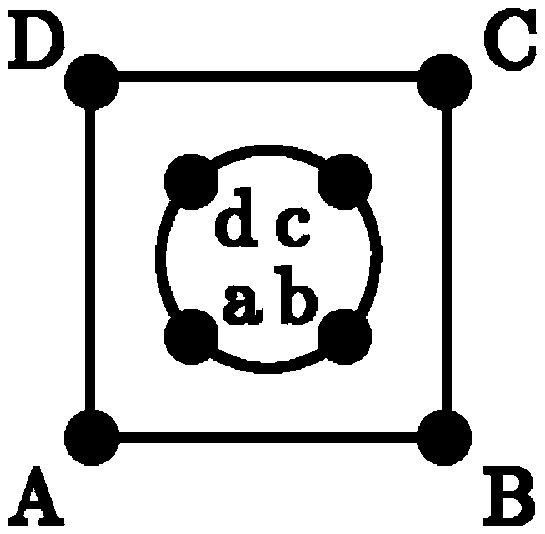}{{\sc van\,der\,Pauw}-{\sc Corbino}-hybrid sample layout}{6}
\vspace*{\fill}
\eject\noindent%
\vspace*{\fill}
\bild{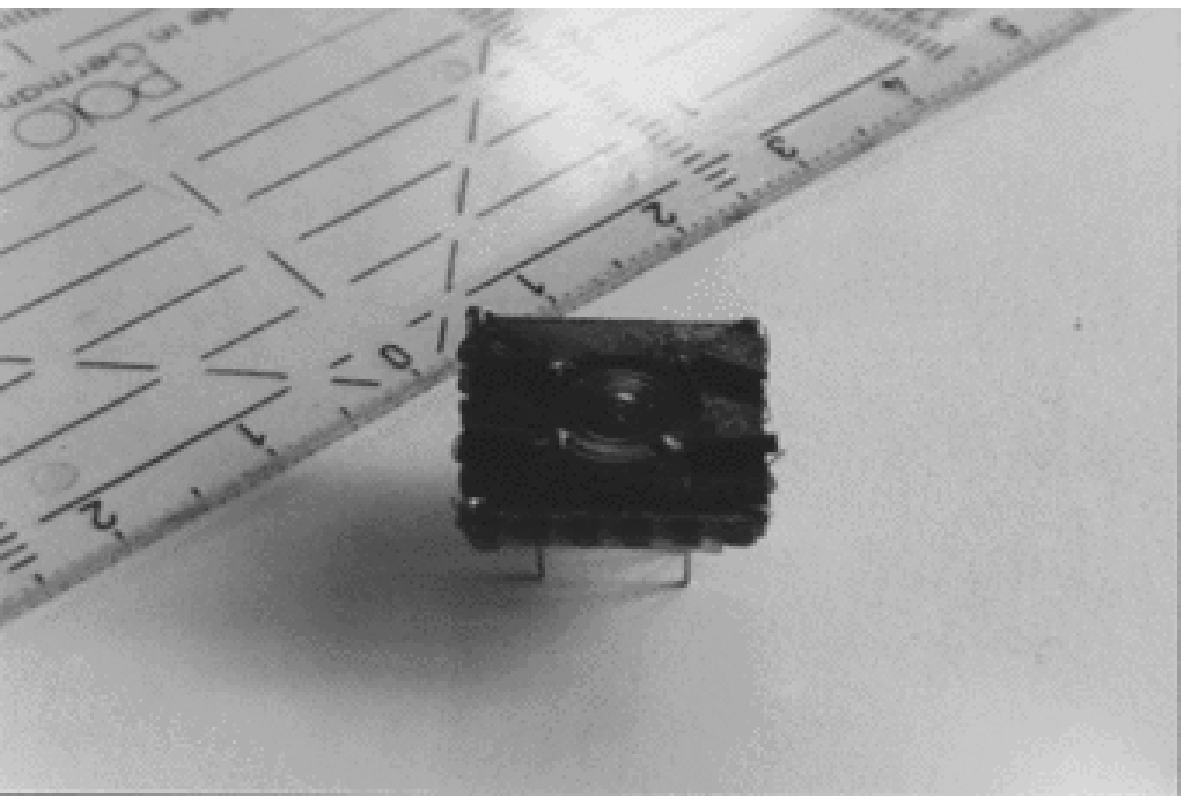}{Photograph of the
                {\sc van\,der\,Pauw}-{\sc Corbino}-hybrid sample}{12}
\vspace*{\fill}
\newpage
%

%
\section{\hspace{5mm}Figures: Experimental results}
\vspace*{\fill}
\bild{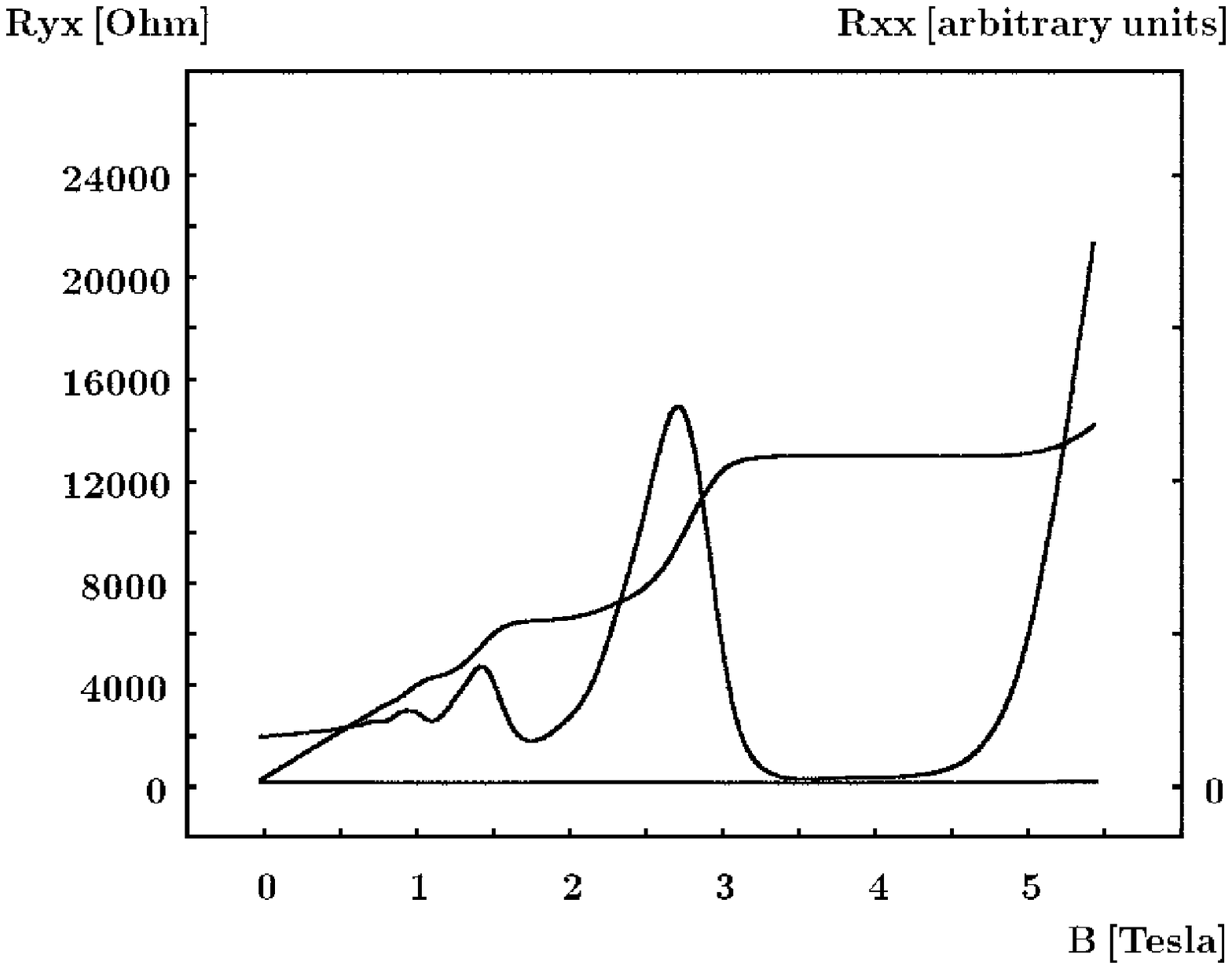}{Measurement 1 on the
                {\sc van\,der\,Pauw}-type
                \lq\lq${\bf centi}$\rq\rq\ sample}{14}
\vspace*{\fill}
\eject\noindent%
\vspace*{\fill}
\bild{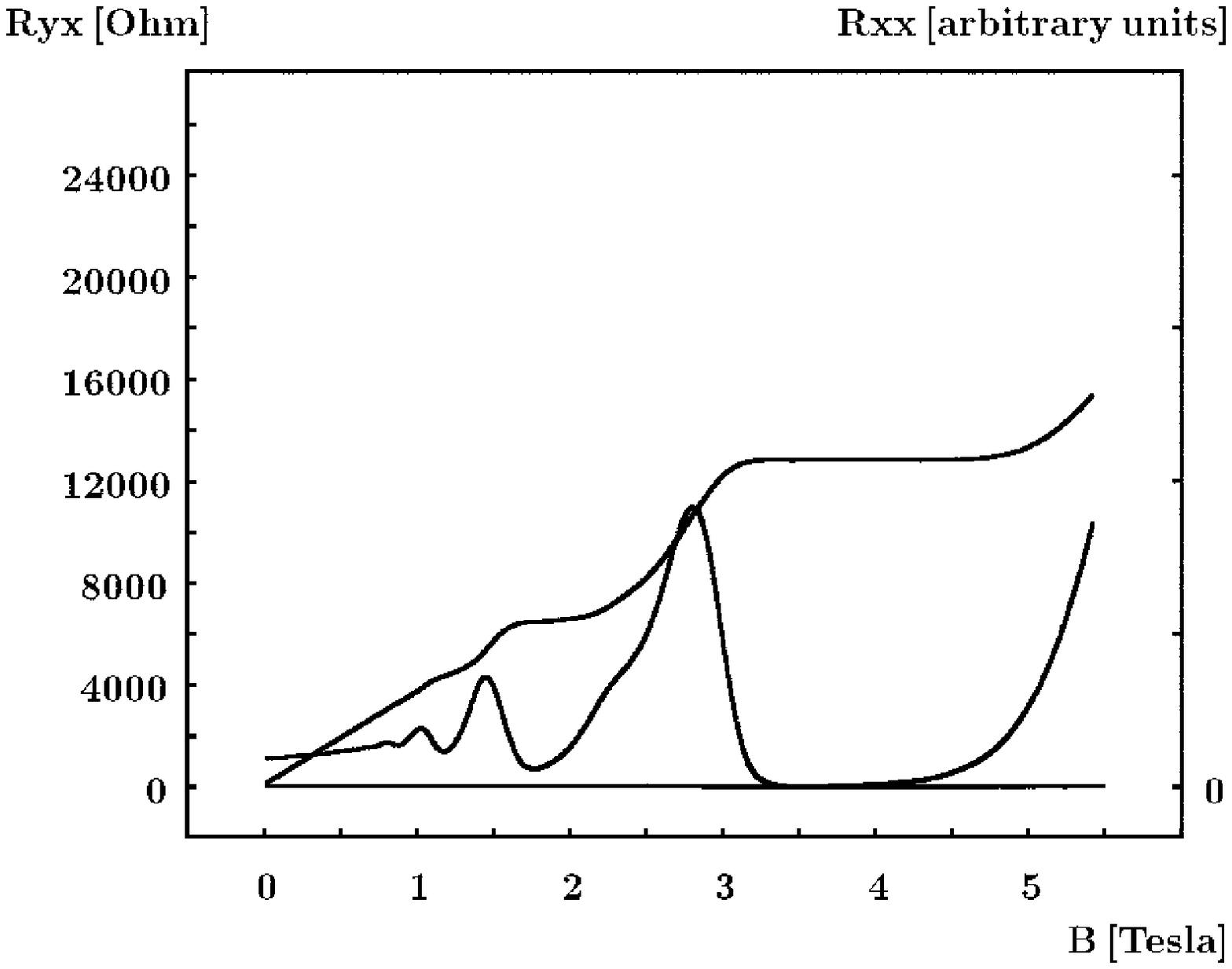}{Measurement 2 on the
                {\sc van\,der\,Pauw}-type
                \lq\lq${\bf centi}$\rq\rq\ sample}{14}
\vspace*{\fill}
\eject\noindent%
\vspace*{\fill}
\bild{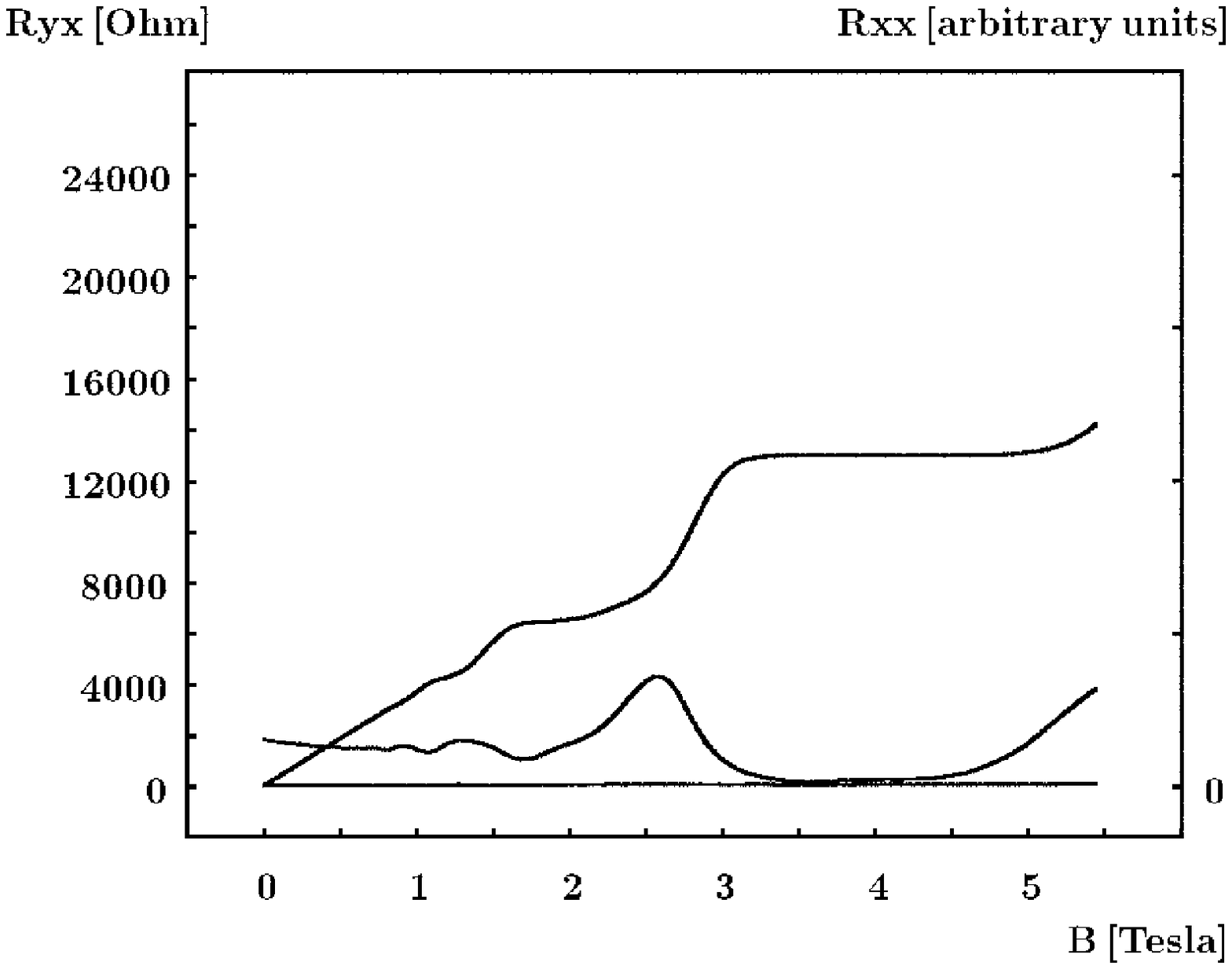}{Measurement 3 on the
                {\sc van\,der\,Pauw}-type
                \lq\lq${\bf centi}$\rq\rq\ sample}{14}
\vspace*{\fill}
\eject\noindent%
\vspace*{\fill}
\bild{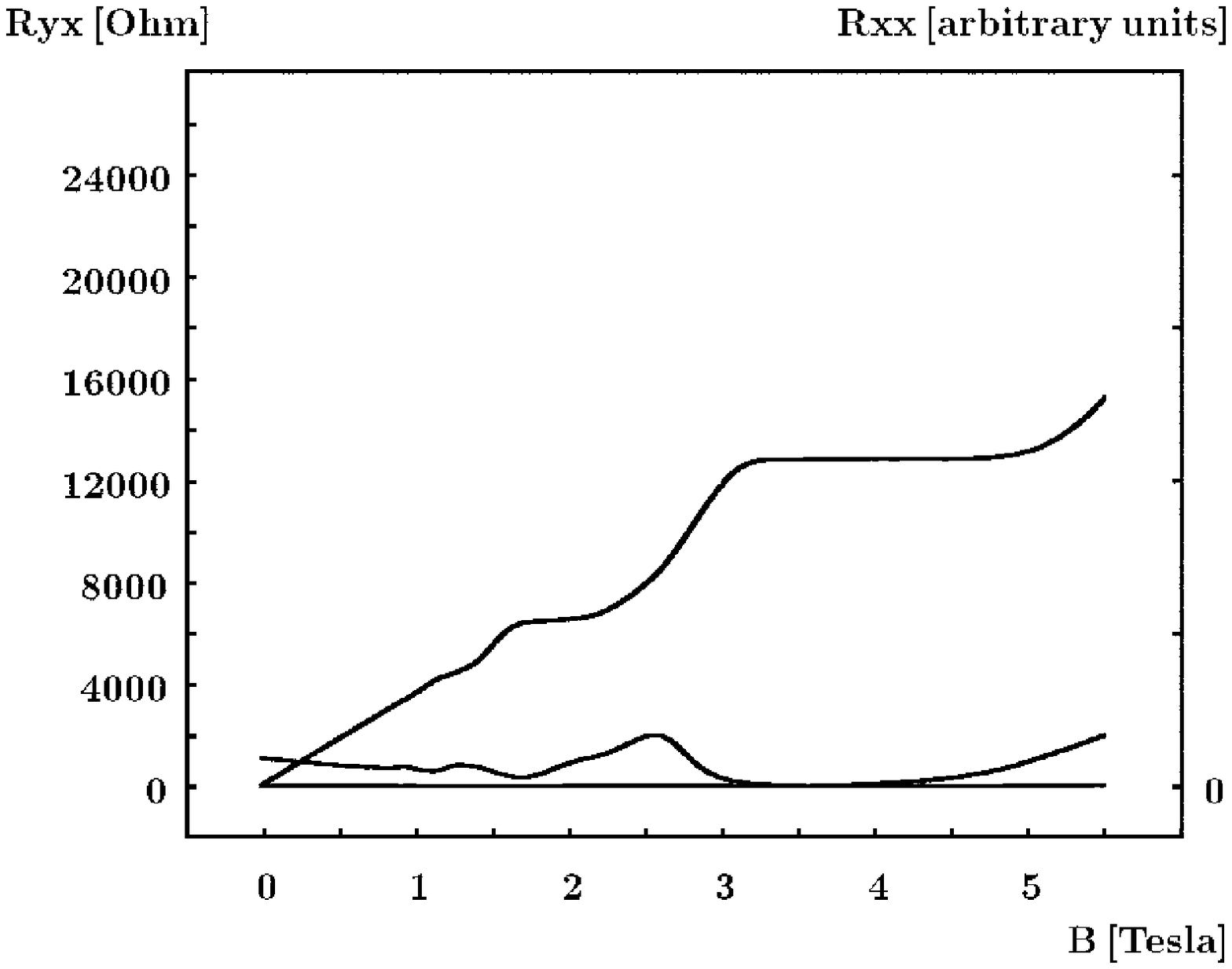}{Measurement 4 on the
                {\sc van\,der\,Pauw}-type
                \lq\lq${\bf centi}$\rq\rq\ sample}{14}
\vspace*{\fill}
\eject\noindent%
\vspace*{\fill}
\bild{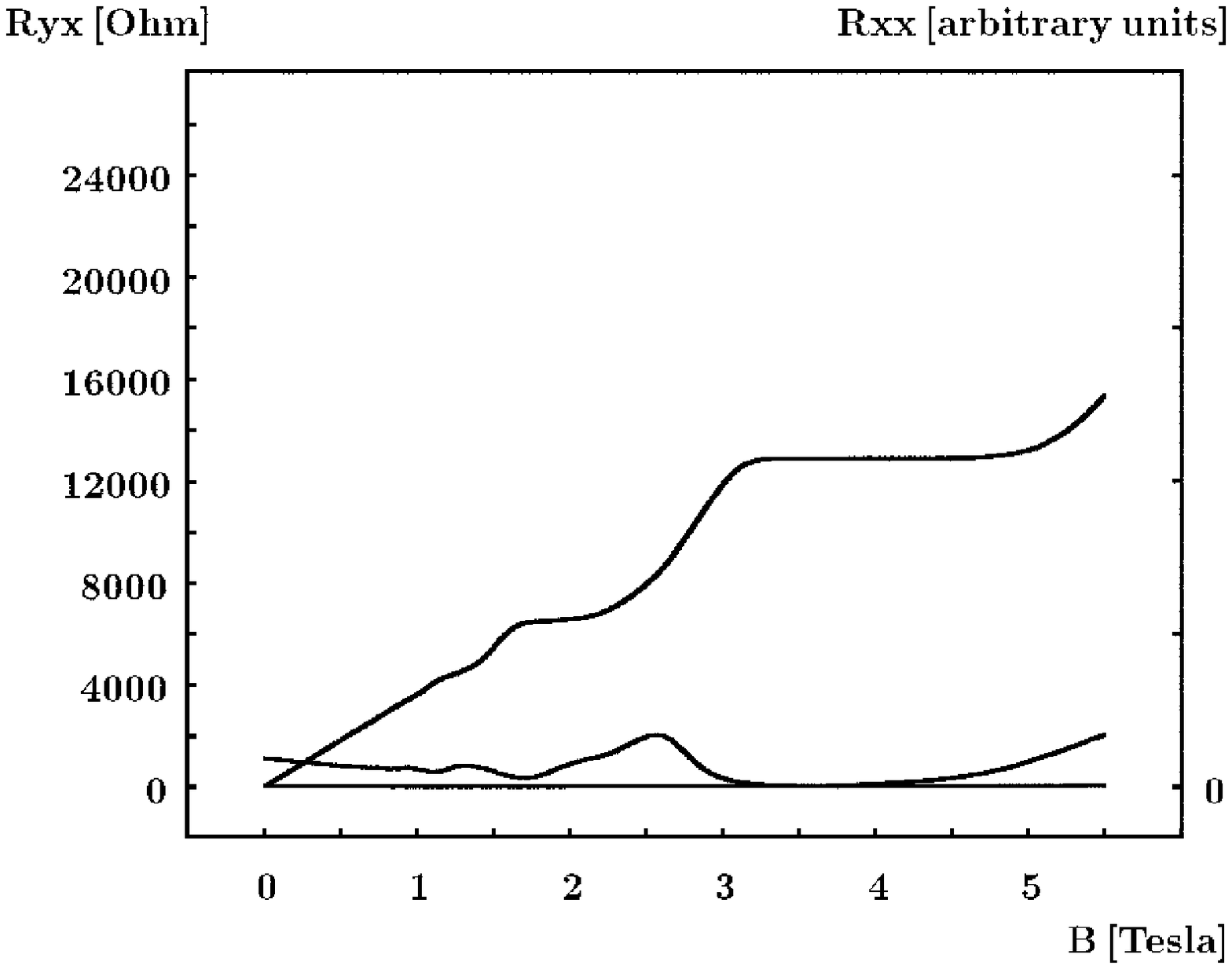}{Measurement 5 on the
                {\sc van\,der\,Pauw}-type
                \lq\lq${\bf centi}$\rq\rq\ sample}{14}
\vspace*{\fill}
\eject\noindent%
\vspace*{\fill}
\bild{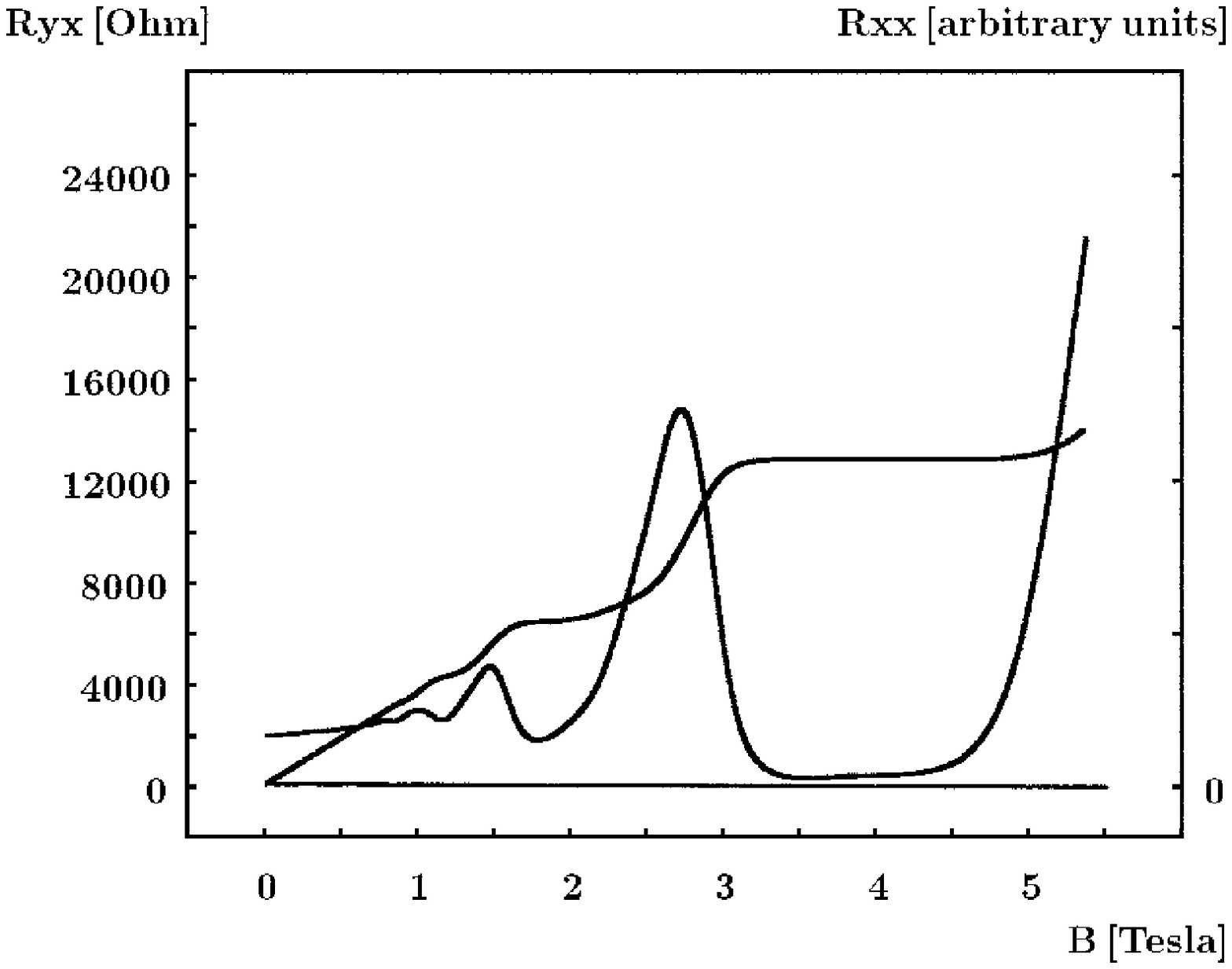}{Measurement 6 on the
                {\sc van\,der\,Pauw}-type
                \lq\lq${\bf centi}$\rq\rq\ sample}{14}
\vspace*{\fill}
\eject\noindent%
\vspace*{\fill}
\bild{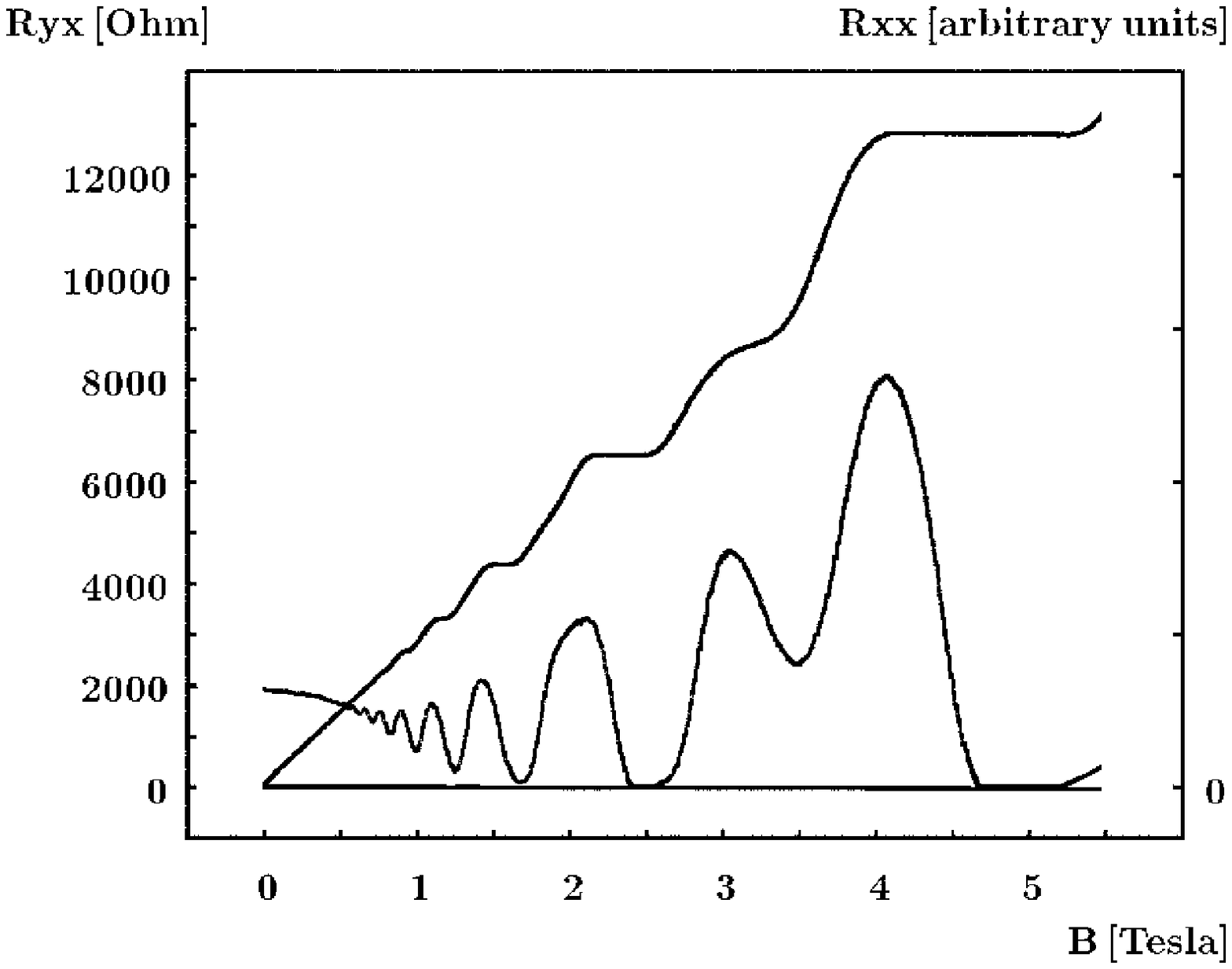}{Measurement on a
                {\sc Hall}-bar-type
                \lq\lq${\bf micro}$\rq\rq\ sample}{14}
\vspace*{\fill}
\eject\noindent%
\vspace*{\fill}
\bild{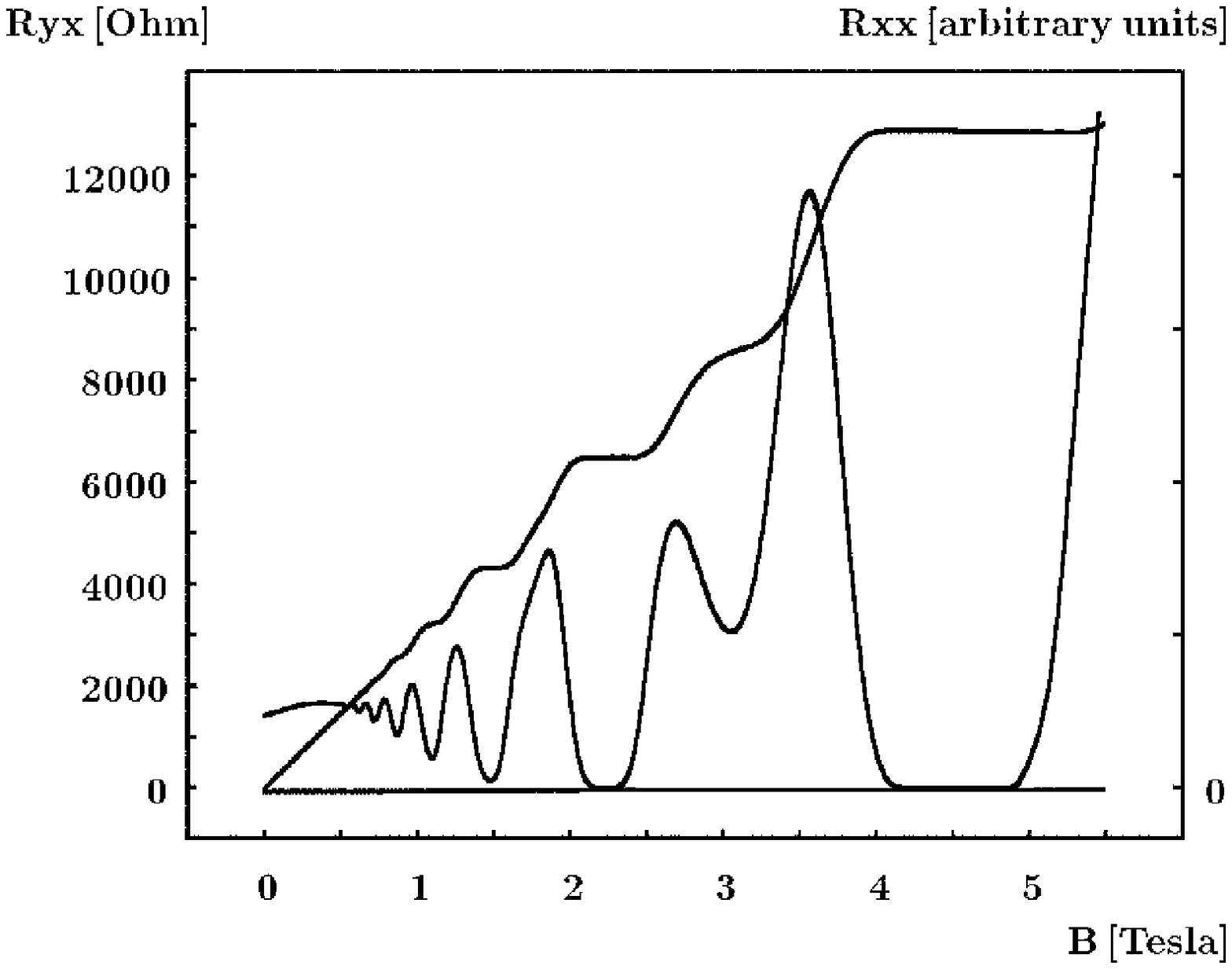}{Measurement on a
                {\sc van\,der\,Pauw}-type
                \lq\lq${\bf milli}$\rq\rq\ sample}{14}
\vspace*{\fill}
\eject\noindent%
\vspace*{\fill}
\bild{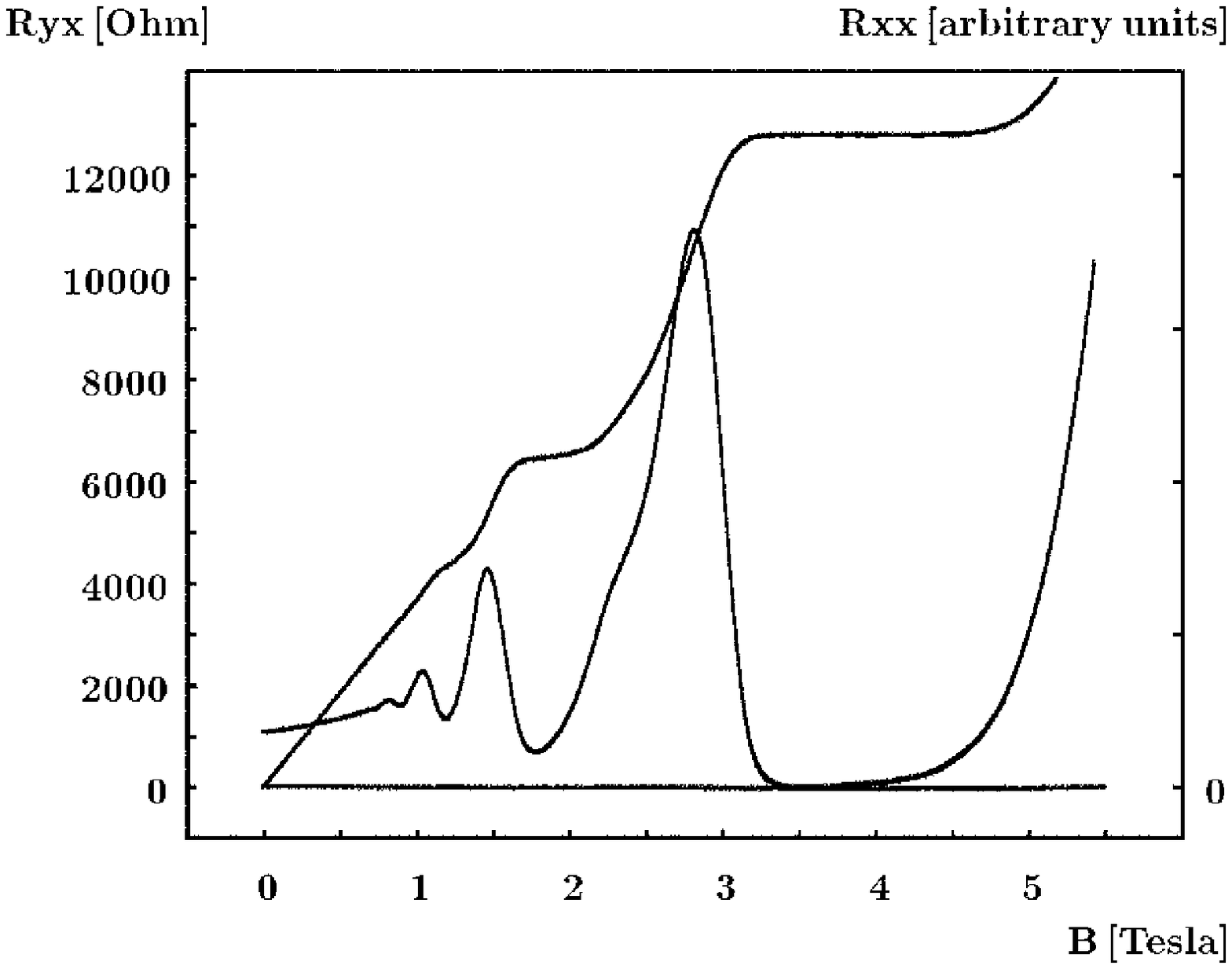}{Measurement on a
                {\sc van\,der\,Pauw}-type
                \lq\lq${\bf centi}$\rq\rq\ sample}{14}
\vspace*{\fill}
\eject\noindent%
\vspace*{\fill}
\bild{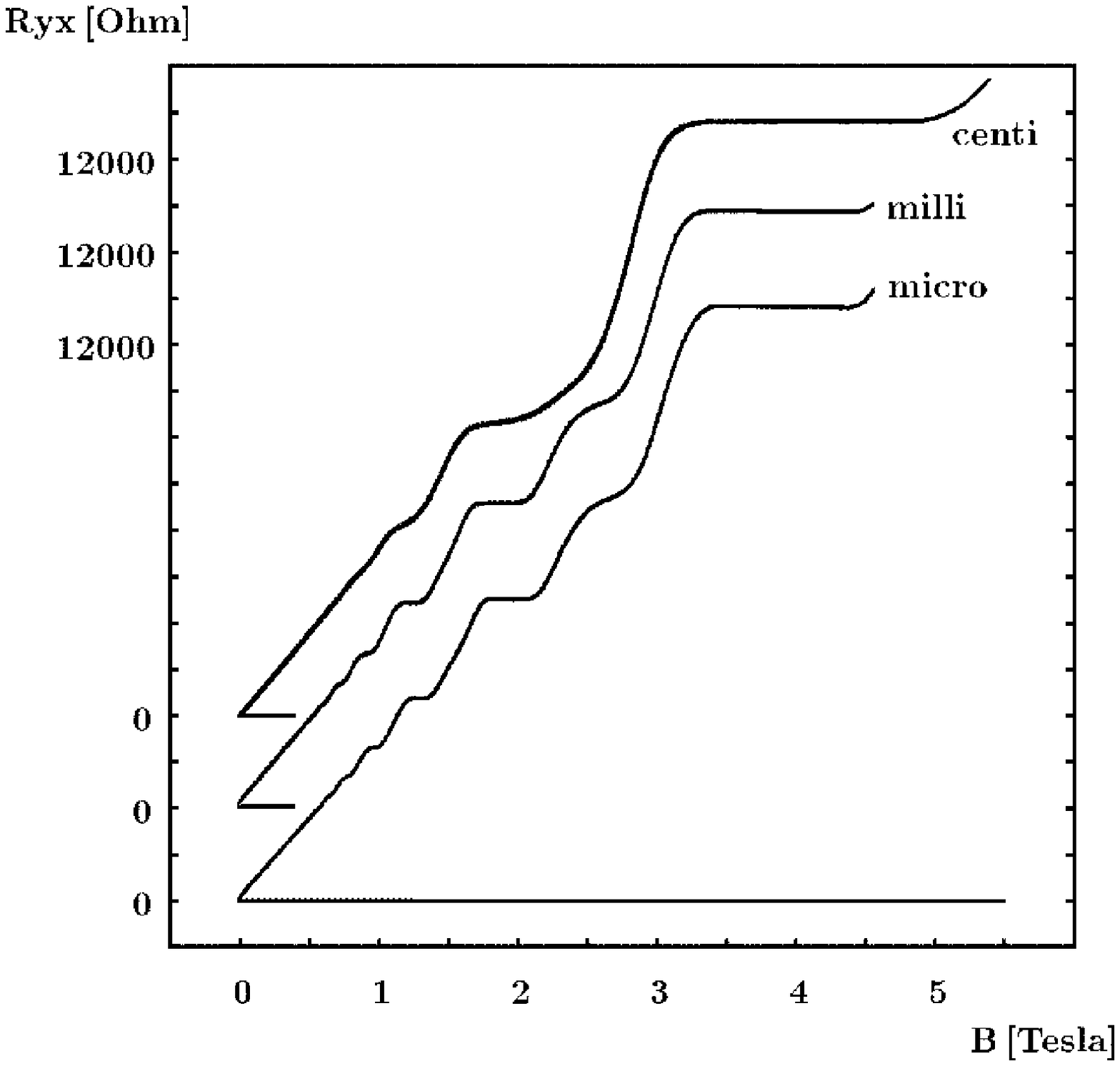}{Scaling in real space:
                Comparison of different sample sizes}{14}
\vspace*{\fill}
\eject\noindent%
\vspace*{\fill}
\bild{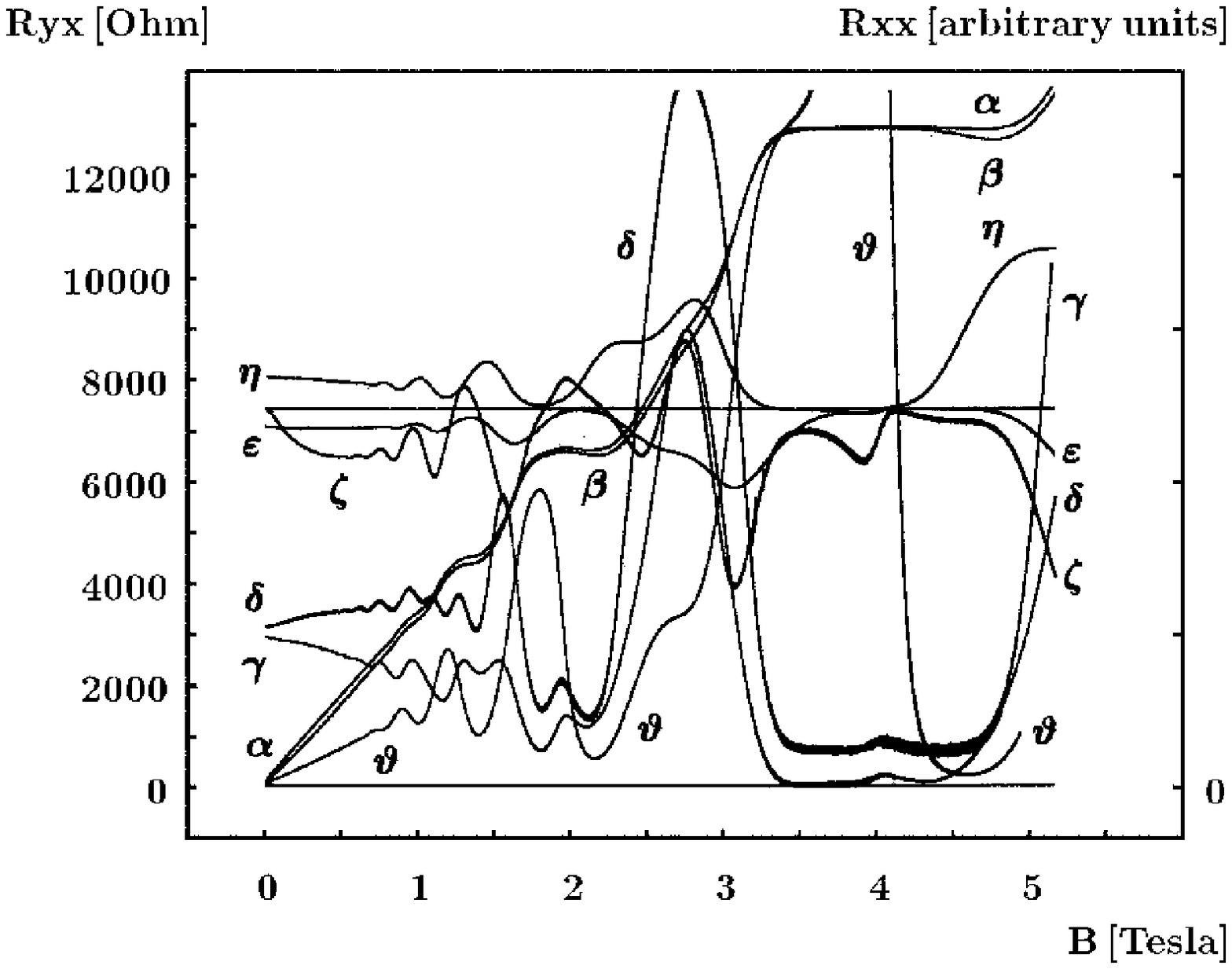}{Measurement on a
               {\sc van\,der\,Pauw}-{\sc Corbino}-hybrid sample}{14}
\vspace*{\fill}
\newpage

%
\section{\hspace{5mm}Tables}
%
%
\begin{table}[h]
\vspace{4cm}
\begin{center}
\begin{tabular}{|r|r|r|r|}
\hline $\nu$ & \lq\lq{\bf micro}\rq\rq\
             & \lq\lq{\bf milli}\rq\rq\
             & \lq\lq{\bf centi}\rq\rq\ \\ \hline
                                           \hline
             {\bf  2} & 100 $\%$ & 121 $\%$ & 179 $\%$ \\ \hline
             {\rm  3} & 100 $\%$ & 150 $\%$ &  50 $\%$ \\ \hline
             {\bf  4} & 100 $\%$ & 117 $\%$ & 150 $\%$ \\ \hline
             {\bf  6} & 100 $\%$ & 117 $\%$ & 100 $\%$ \\ \hline
             {\bf  8} & 100 $\%$ & 117 $\%$ &  50 $\%$ \\ \hline
             {\bf 10} & 100 $\%$ & 100 $\%$ &   0 $\%$ \\ \hline
             {\bf 12} & 100 $\%$ & 100 $\%$ &   0 $\%$ \\ \hline
\end{tabular}\normalsize
\end{center}
\vspace{0.5cm}
\caption{Relative plateau width in Fig.\,26}
\end{table}
\eject\noindent%
%
%
\begin{table}[h]
\vspace{4cm}
\begin{center}
\begin{tabular}{|c|c|c|c|c|c|}
\hline meas.\ no.\ & time  & $I_{in}$  & $U_{out}$ & $U_{out}$ range \\ \hline
                                                                        \hline
                1   & 00:00 & AC        & BD        & 3 mV           \\
                    &       & AB        & CD        & 0.3 mV         \\ \hline
                2   & 00:20 & BD        & CA        & 3 mV           \\
                    &       & BC        & AD        & 0.3 mV         \\ \hline
                3   & 00:37 & CA        & DB        & 3 mV           \\
                    &       & CD        & AB        & 0.3 mV         \\ \hline
                4   & 00:45 & DB        & AC        & 3 mV           \\
                    &       & DA        & BC        & 0.3 mV         \\ \hline
                5   & 00:60 & AC        & BD        & 3 mV           \\
                    &       & AB        & CD        & 0.3 mV         \\ \hline
                6   & 01:31 & AC        & BD        & 3 mV           \\
                    &       & AB        & CD        & 0.3 mV         \\ \hline
\end{tabular}\normalsize
\end{center}
\vspace{0.5cm}
\caption{Measurements depicted in Fig.\,17 - Fig.\,22}
\end{table}
\eject\noindent%
%
%
\begin{table}[h]
\vspace{5cm}
\begin{center}
\begin{tabular}{|c|c|c|c|c|c|}
\hline curve               & $I_{in}$  & $U_{out}$ & $U_{out}$ range
              & zero line & overall characteristics \\ \hline\hline
       \boldmath$\alpha$   & AC & DB & 3 mV
              & low
              & standard transversal curve
              \\ \hline
       \boldmath$\beta$    & ac & db & 3 mV
              & low
              & standard transversal curve
              \\ \hline
       \boldmath$\gamma$   & AB & CD & 0.3 mV
              & low
              & longitudinal curve with neg.\ slope
              \\ \hline
       \boldmath$\delta$   & ab & cd & 0.3 mV
              & low
              & longitudinal curve with pos.\ slope
              \\ \hline
       \boldmath$\epsilon$ & AD & bd & 3 mV
              & mid
              & longitudinal curve with pos.\ slope
              \\ \hline
       \boldmath$\zeta$    & AC & bd & 0.3 mV
              & mid
              & curve with second derivative content
              \\ \hline
       \boldmath$\eta$     & AB & bd & 3 mV
              & mid
              & longitudinal curve with neg.\ slope
              \\ \hline
       \boldmath$\vartheta$   & AC & bB & 3 mV
              & low
              & hybrid curve
              \\ \hline
\end{tabular}
\end{center}
\vspace{0.5cm}
\caption{Evaluation of Fig.\,27}
\end{table}
\eject
%

%
\end{document}